\documentclass[12pt]{article}
\usepackage{amssymb, amsmath, amsthm}
\usepackage{graphicx}
\usepackage{fullpage,booktabs}
\usepackage{siunitx}
\usepackage{subcaption}
\usepackage{multirow}
\usepackage{multirow}
\usepackage{bm}
\usepackage{dsfont}
\usepackage{natbib}
\newcommand{\vertiii}[1]{{\|\kern-0.2ex| #1 
    \|\kern-0.2ex|}}

\newcommand{\argmin}{\operatorname*{arg \ min}}

\newcommand{\mbX}{\boldsymbol{X}}

\newcommand{\mbW}{\boldsymbol{W}}

\usepackage{etoolbox}
\AtBeginEnvironment{algorithm}{\setstretch{1.35}}

\newcommand{\blA}{\boldsymbol{A}}

\newcommand{\blE}{\boldsymbol{E}}

\newcommand{\blH}{\boldsymbol{H}}

\newcommand{\blJ}{\boldsymbol{J}}

\newcommand{\blP}{\boldsymbol{P}}

\newcommand{\blU}{\boldsymbol{U}}
\newcommand{\blV}{\boldsymbol{V}}

\newcommand{\blX}{\boldsymbol{X}}
\newcommand{\blY}{\boldsymbol{Y}}
\newcommand{\blZ}{\boldsymbol{Z}}

\newcommand{\ble}{\boldsymbol{e}}

\newcommand{\blj}{\boldsymbol{j}}
\newcommand{\blk}{\boldsymbol{k}}

\newcommand{\blu}{\boldsymbol{u}}
\newcommand{\blv}{\boldsymbol{v}}
\newcommand{\blw}{\boldsymbol{w}}
\newcommand{\blx}{\boldsymbol{x}}
\newcommand{\bly}{\boldsymbol{y}}
\newcommand{\blz}{\boldsymbol{z}}

\newcommand{\bgalpha}{\boldsymbol{\alpha}}
\newcommand{\bgbeta}{\boldsymbol{\beta}}

\newcommand{\bgzeta}{\boldsymbol{\zeta}}
\newcommand{\bgeta}{\boldsymbol{\eta}}
\newcommand{\bgtheta}{\boldsymbol{\theta}}

\newcommand{\bgmu}{\boldsymbol{\mu}}

\newcommand{\bgpi}{\boldsymbol{\pi}}

\newcommand{\bgLambda}{\boldsymbol{\Lambda}}

\usepackage{hyperref}
\usepackage[table]{xcolor}
\newcolumntype{L}{>$l<$}
\theoremstyle{plain}

\newtheorem{example}{Example}

\newtheorem{lemma}{Lemma}
\newtheorem{theorem}{Theorem}

\newtheorem{remark}{Remark}
\newtheorem{assumption}{Assumption}
\newtheorem{corol}{Corollary}

 \newcommand{\indep}{\perp\!\!\!\!\perp}

\newcommand{\innerpoduct}[2]{\left\langle #1, #2\right\rangle }
\newcommand{\norm}[1]{\left\lVert#1\right\rVert}

\usepackage{tocloft}

\begin{document}

\title{\textbf{Subspace decompositions for association structure learning in multivariate categorical response regression}}
\author{Hongru Zhao$^\dagger$, Aaron J. Molstad$^{\dagger,\star}$, and Adam J. Rothman$^\dagger$\\
School of Statistics, University of Minnesota, Minneapolis, MN$^\dagger$\\
Department of Statistics, University of Florida, Gainesville, FL$^\star$
}
\date{}
\maketitle

\begin{abstract}
Modeling the complex relationships between multiple categorical response variables as a function of predictors is a fundamental task in the analysis of categorical data. However, existing methods can be difficult to interpret and may lack flexibility. To address these challenges, we introduce a penalized likelihood method for multivariate categorical response regression that relies on a novel subspace decomposition to parameterize interpretable association structures. Our approach models the relationships between categorical responses by identifying mutual, joint, and conditionally independent associations, which yields a linear problem within a tensor product space. We establish theoretical guarantees for our estimator, including error bounds in high-dimensional settings, and demonstrate the method's interpretability and prediction accuracy through comprehensive simulation studies.
  \end{abstract}
  
\def\spacingset#1{\renewcommand{\baselinestretch}%
{#1}\small\normalsize} \spacingset{1.25}

\section{Introduction}
We consider a multivariate response regression where each of the response variables is categorical. Specifically, let $\blX \in \mathcal{X} \subseteq \mathbb{R}^p$ be the predictor vector and let $\blZ = (Z_1,\cdots,Z_q)^\top$ be the multivariate categorical response. The $k$th component of the response, $Z_k$, has $J_k$ numerically coded outcome categories with $J_k \geq 2$ for  $k\in[q]$, where $[m]$ is defined as $\{1,\ldots, m\}$ for positive integer $m$. The essential problem is to model the conditional distribution $\blZ|\blX=\blx$ whose joint probability mass function is given by
\begin{equation}\label{equ:prob_dist_pi}
\pi_{\blj}(\blx) := \mathbb{P}( Z_1=j_1, \cdots,Z_q=j_q \big\vert \blX=\blx )\geq 0,
\end{equation}
for any $\blj=(j_1,\cdots,j_q)\in \blJ:=[J_1]\times \cdots \times [J_q]$, where $\ j_l\in [J_l]$ for all $l \in [q]$. For a given $\blx$, $\blZ$ has a multivariate version of the single-trial multinomial distribution. If, for a given $\blx$, one were to observe $v \geq 1$ independent realizations of $\blZ$, say $\blz_1, \dots, \blz_v$, then the probability mass function corresponding to \eqref{equ:prob_dist_pi} would be given by 
\begin{equation*}
    \frac{v!}{ \prod_{ \blj\in \blJ } y_{\blj}! } \prod_{ \blj \in \blJ } \left\{\pi_{\blj }(\blx) \right\}^{y_{\blj}},
\end{equation*}
where $y_{\blj} := \sum_{i=1}^v \mathbf{1}(\blz_i = \blj)$ for each $\blj \in \blJ$. 

For a given $\blx$, if $v$ is sufficiently large, one could model \eqref{equ:prob_dist_pi} using standard methods for the analysis of $q$-way contingency tables, a classical problem in categorical data analysis \citep{mccullagh1989generalized,christensen1997log,agresti2002categorical}. However, when one needs to model \eqref{equ:prob_dist_pi} for all $\blx \in \mathcal{X}$, methods for contingency tables cannot be applied. For example, in many applications, for every subject in the study we observe (or impose) a distinct $\blx$, and observe the outcome of only a single trial, $v = 1$.  Instead, one could model \eqref{equ:prob_dist_pi} using existing methods for multinomial regression (or in statistical learning terminology, multiclass classification). Notice that \eqref{equ:prob_dist_pi} could be equivalently defined in terms of a ``univariate'' categorical response variable, $\blZ^\star$, with $|\blJ|$ many outcome categories: one corresponding to each distinct element of $\blJ$. Letting $f:\blJ \to [|\blJ|]$ be any bijective function, it would thus be natural to model $\mathbb{P}(\blZ^\star = f(\blj) \mid \blX = \blx) = \pi_{\blj}(\blx) = \mathbb{P}( Z_1=j_1, \cdots,Z_q=j_q \big\vert \blX=\blx)$ using multinomial logistic regression \citep{agresti2002categorical,vincent2014sparse}; linear or quadratic discriminant analysis \citep{hastie2009elements,mai2019multiclass}; or nonparametric methods. Modeling the conditional distribution $\blZ^\star \mid \blX$ using one of these methods is appealing because they allow for arbitrary dependence among the $q$ categorical response variables. 

However, off-the-shelf application of methods designed for a univariate categorical response may be problematic. In particular, these methods would fail to exploit that $\blZ^\star$ is constructed from $q$ distinct response variables. This negatively affects both estimation efficiency and interpretability of the fitted model. Moreover, for even moderate $q$, the cardinality of $\blJ$, $|\blJ|$, will be large. As a consequence, with small sample sizes, many outcome category combinations $\blj$ will not be observed in the training data. If one used a multinomial logistic regression in this situation, the maximum likelihood estimator would not exist.  In this work, we propose a new method for fitting \eqref{equ:prob_dist_pi} that allows practitioners to discover parsimonious and interpretable dependence structures amongst responses. 

To motivate our approach, consider a multinomial logistic regression model for \eqref{equ:prob_dist_pi} with $\blx \in \mathbb{R}$ (i.e., $p=1$), 
\begin{equation}\label{eq:multinom_example}
    \pi_{ \blj } ( \blx ) = \mathbb{P}\left( Z_1=j_1,\dots, Z_q=j_q |\blX=\blx \right) =  \frac{\exp(\blx\cdot \bgzeta_{\blj} )}{\sum_{\blj\in \blJ } \exp(\blx \cdot \bgzeta_{\blj} ) },~~~ \blj \in \blJ, ~~~ \sum_{\blj \in \blJ} \bgzeta_{\blj} = 0,
\end{equation}
where $\bgzeta=\{\bgzeta_{\blj}\}_{\blj\in \blJ}$ is an unknown tensor. In full generality, $\bgzeta \in \{\blv \in \mathbb{R}^{[J_1] \times \cdots \times [J_q]}: \sum_{\blj \in \blJ} \blv_{\blj} = 0\}$, which implies no restrictions on the dependence amongst responses: their dependence can be arbitrarily complex. Restrictions on the dependence between responses under \eqref{equ:prob_dist_pi} can often be represented as constraints on the space of the coefficients $\bgzeta$.  For example, in the case that $q = 2$, $J_1 = J_2 = 2$, if $\bgzeta \in \mathfrak{C}^0$, where
$$\mathfrak{C}^0 = \{\bgzeta \in \mathbb{R}^{[2] \times [2]}: \bgzeta_{(1,1)} + \bgzeta_{(2,2)} - \bgzeta_{(1,2)} - \bgzeta_{(2,1)} = 0 \},$$
then $Z_1 \indep Z_2 \mid \blX$. Intuitively, $\mathfrak{C}^0$ is the set of coefficients for which the log odds ratio between the two responses is zero for all $\blx$. 
This observation motivated \citep{molstad2023likelihood} to propose a regularized maximum likelihood estimator of $\bgzeta$ that shrinks coefficients towards the set $\mathfrak{C}^0$. For applications with $J_l \geq 2$ and $q \leq 3$, \citep{molstad2023likelihood} generalized the set $\mathfrak{C}^0$ to correspond to coefficients with all local log odds ratios equal to zero. Their approach thus allowed practitioners to discover only whether responses are mutually independent ($\bgzeta \in \mathfrak{C}^0$) or are arbitrarily dependent ($\bgzeta \not\in \mathfrak{C}^0$). When $q \geq 3$, however, there are many other parsimonious dependence structures which are ``intermediate" to mutual independence and arbitrary depedence. In this work, we generalize the approach of \citep{molstad2023likelihood}, allowing practicioners to discover much more complex, interpretable dependence structures.

As we just described, to learn the association structure for \eqref{equ:prob_dist_pi}, it is crucial to identify whether the regression coefficients reside within a specific subspace. Representing the linear subspace \(\mathfrak{L}\) of \(\mathbb{R}^k\) can be approached in two ways: external and internal. For the external representation, consider \(\mathfrak{L} = \ker(\boldsymbol{A}) = \{\boldsymbol{v} \in \mathbb{R}^k; \boldsymbol{A} \boldsymbol{v} = \boldsymbol{0}\}\) for some matrix \(\boldsymbol{A}\). Then regularizing \(\boldsymbol{v}\) towards the subspace \(\ker(\boldsymbol{A})\) can be achieved by penalizing the term \(\|\boldsymbol{A} \boldsymbol{v}\|_2\). In this sense, \cite{molstad2023likelihood} achieve structure learning via an external subspace representation. In contrast, for the internal representation, we can set \(\mathfrak{L} = \operatorname{span}\big(\{\boldsymbol{e}_1, \ldots, \boldsymbol{e}_s\}\big)\) and $\blv=v_1 \ble_1 + v_2 \ble_2+ \cdots + v_k \ble_k $, where \(\boldsymbol{e}_1, \ldots, \boldsymbol{e}_k\) form an orthonormal basis for \(\mathbb{R}^k\). Then regularizing $\blv$ towards the subspace $\operatorname{span}\big(\{\boldsymbol{e}_1, \ldots, \boldsymbol{e}_s\}\big)$ can be achieved by penalizing the terms $ v_{s+1},\cdots,v_{k}$. This is the approach we take in this paper, by selecting an orthonormal basis and penalizing the coordinates to achieve association structure learning. 

For example, if $J_1=J_2=2$, we can define 
\begin{equation*}
    \bgzeta^*= {v_{11}}  g_1g_1^\top+ v_{12} g_1g_2^\top + v_{21} g_2g_1^\top  + v_{22} g_2g_2^\top,
\end{equation*}
where $g_i=\frac{1}{\sqrt{2}}\big(1, (-1)^{i+1}\big)^\top,i\in\{1,2\}$. Here, $g_1g_1^\top$ represents the overall effect, $g_1g_2^\top$ denotes the main effect of category 1, $g_2g_1^\top$ is for the main effect of category 2, and $g_2g_2^\top$ captures the interaction effect between categories 1 and 2. 
Because $\mathfrak{C}^0=\operatorname{span}\{g_1g_1^\top,g_1g_2^\top,g_2g_1^\top\}$, if $v_{22} = 0$, then $\bgzeta^* \in \mathfrak{C}^0$. That is, by carefully constructing the internal subspace representation, sparsity in the corresponding coefficients can imply parsimonious association structures among responses.  This observation is central to our methodological developments, and one of our main contributions is the explicit construction of a flexible, interpretable internal subspace representation. 

Multivariate categorical response regression without predictors serves as an extension of contingency table analysis, allowing for a more comprehensive examination of categorical variable interrelations. The Poisson log-linear model is used for association structure modeling of multiple categorical responses without predictors, with the connection between log-linear models for frequencies and multinomial response models for proportions being extensively studied \citep{mccullagh1989generalized,christensen1997log}.


In this paper, we will study structure learning via an internal subspace representation. We present a reparameterization via subspace decomposition and obtain a unifying framework for both multinomial and Poisson categorical response regression models in high dimensions. Complex dependencies between response variables can be systematically modeled, encompassing all possible association structures, including mutual independence, joint independence, and conditional independence among response variables. We apply group lasso penalty \cite{yuan2006model} and overlapping group lasso \citep{zhao2009composite,jenatton2011proximal} over reparameterization parameters. We apply the accelerated proximal gradient descent algorithm to solve the convex optimization problem. We prove an
error bound that illustrates our estimator’s performance in high-dimensional settings. A key theoretical advancement in our research is the derivation of restricted strong convexity conditions specific to multivariate categorical response regression, which notably incorporates the Rademacher complexity associated with general norm penalties. Finally, simulation studies validate our method's effectiveness in terms of interpretability, and prediction accuracy.
 
We conclude this section by introducing notation to be used for the remainder of the article. First, let $\lambda_{\max}(\blA)$ and $\lambda_{\min}(\blA)$ denote the maximum and minimum eigenvalues of the real symmetric matrix $\blA$. 
For any vector (resp. matrix) $\blX$, define the Euclidean (resp. Frobenius) norm $\norm{\blX}=\sqrt{ \operatorname{tr}(\blX^\top \blX)}$. Let \( \mathbf{1}_m \) denote a vector of ones of length \( m \). For matrices $\blX$ and $\blY$ of the same size, define the Frobenius inner product $\innerpoduct{\blX}{\blY}=\operatorname{tr}(\blX^\top\blY)$ and define the operator norm $\norm{\blX}_{{\rm op}}=\sqrt{\lambda_{\max}(\blX^\top\blX)}$. Define the maximum norm $\norm{ \blX }_{\infty}=\max_{i,j}|x_{ij}|,\blX=\{x_{i,j}\}_{1\leq i\leq n,1\leq j\leq m}$. Let $\norm{\bgbeta}_0 = \sum_{j,k} \mathbf{1}(\bgbeta_{j,k} \neq 0)$ for matrix $\bgbeta$. 
Let $I$ be the identity matrix. Let \( \bm{1}_m \) denote a vector of ones of length \( m \).
When $\blX$ and $\blY$ are matrices, let $\blX\otimes \blY$ denote the Kronecker product between $\blX$ and $\blY$. {When $\blU$ and $\blV$ are vector spaces, let $\blU \otimes \blV$ be the tensor product of $\blU$ and $\blV$.} Finally, let $\otimes_t(\blu,\blv)$ denote the tensor product between vector $\blu$ and $\blv$.

\section{Association structure learning via subspace decomposition}\label{sec:2subspace_learning}
\subsection{Overview}


Assume the response has $q \geq 2$ categorical components with $J_1,\cdots,J_q$ categories, respectively. 
Define the Cartesian product of an indexed family of sets $\blJ=[J_1]\times [J_2]\times \cdots \times [J_q]$. The cardinally of set $\blJ$ is $|\blJ|=\prod_{i=1}^q J_i$. {Let $\mathbb{R}^{\blJ}$ and $\mathbb{N}^{\blJ}$ denote the spaces of $\blJ$ arrays with entries that are real numbers and whole numbers, respectively. That is, $\bly=\{y_{\blj}\}_{\blj\in \blJ} \in \mathbb{F}^{\blJ}$ if and only if $y_{\blj}\in \mathbb{F}$ for any $\blj \in \blJ$, where field $\mathbb{F}$ can take $\mathbb{R}$ and $\mathbb{N}$.} For a $q$-way array of shape $\blJ$, let $\bly^{\blJ}=\{y_{  \blj }\}_{\blj\in \blJ}\in \mathbb{R}^{\blJ}$, where $\blj=(j_1,\cdots,j_q), j_1\in [J_1],\cdots, j_q\in [J_q]$. Define the $\blJ$-vectorization of $\bly^{\blJ}$ as
\begin{equation}\label{def:vec_J}
    \operatorname{vec}_{\blJ} (\bly^{\blJ}):=(y_{1,1,\cdots,1},y_{2,1,\cdots,1},\cdots,y_{J_1,1,\cdots,1},y_{1,2,\cdots,1},\cdots,  y_{J_1,J_2,\cdots,J_q})^\top \in \mathbb{R}^{|\blJ|}.
\end{equation}
Define the inverse $\blJ$-vectorization for vector $\operatorname{vec}_{\blJ} (\bly^{\blJ})$ as \( \operatorname{vec}^{-1}_{\blJ}(\operatorname{vec}_{\blJ} (\bly^{\blJ}))=\bly^{\blJ}.\)

Let the sample data from the $i$th observational unit be denoted $(\blx_i,\bly^{\blJ}_i)$, where $\blx_i\in \mathbb{R}^p$ and $\bly^{\blJ}_i\in \mathbb{N}^{\blJ}$, and let $\bly_i=\operatorname{vec}_{\blJ}(\bly_i^{\blJ})\in \mathbb{N}^{|\blJ|}$. We assume that $\bly_1^{\blJ},\cdots,\bly_n^{\blJ}$ are independent. Similar to \cite{molstad2023likelihood}, we use a multinomial logistic regression model for the $q$ response variables. Specifically, we assume that $\bly_i$ is a realization of a multinomial random vector with 
index $n_i \geq 1$ and category probabilities 
\begin{equation}\label{def:multinomial_prob}
    \operatorname{vec}_{\blJ}\{\bgpi^{\blJ}(\blx_i)\}=\frac{e^{\bgtheta \blx_i }}{\innerpoduct{\bm{1}_{|\blJ|}}{ e^{\bgtheta \blx_i } }} \in \mathbb{R}^{|\blJ|},
\end{equation}
where $\bgtheta\in \mathbb{R}^{|\blJ| \times p}$. In certain settings, we may treat the elements of $\bly_i$ as independent Poisson random variables with mean 
\begin{equation}\label{def:poisson_mean}
    \operatorname{vec}_{\blJ}\{\bgmu^{\blJ}(\blx_i)\}=e^{\bgtheta \blx_i } \in \mathbb{R}^{|\blJ|},
\end{equation}
and category probabilities $\operatorname{vec}_{\blJ}\{\bgpi^{\blJ}(\blx_i)\} = e^{\bgtheta \blx_i }\{\innerpoduct{\bm{1}_{|\blJ|} }{e^{\bgtheta \blx_i }}\}^{-1},$
where 
$\bgmu^{\blJ}(\blx)=\{\mu_{\blj}(\blx)\}_{\blj\in \blJ}$ and $\bgpi^{\blJ}(\blx)=\{\pi_{\blj}(\blx)\}_{\blj\in \blJ}$ are $q$-dimensional arrays.

The $i$th observational unit's contribution to the negative log-likelihood of multinomial and Poisson categorical response models are
\begin{equation}\label{equ:Multivariate_loss}
    \ell_{\rm Mult}(\bgtheta \blx_i,\bly_i ) =-\innerpoduct{\bly_i}{\bgtheta \blx_i} + n_i\cdot \log\big( \innerpoduct{\bm{1}_{|\blJ|}}{e^{\bgtheta \blx_i}}\big),
    \qquad
n_i=\innerpoduct{\bm{1}_{|\blJ| }}{ \bly_i},
\end{equation}
and
\begin{equation}\label{equ:Poisson_loss}
    \ell_{\rm Pois}(\bgtheta \blx_i,\bly_i )=-\innerpoduct{\bly_i}{\bgtheta \blx_i}+ \innerpoduct{\bm{1}_{|\blJ|} }{e^{\bgtheta \blx_i}}.
\end{equation}
The function in \eqref{equ:Multivariate_loss} is sometimes referred to as cross-entropy loss.

When considering the multinomial categorical response model, we impose the constraint $ { \bm{1}_{ |\blJ| }^\top }{ \bgtheta }=\bm{0}$ to address the identifiability issue. Define the linear subspace $\mathcal{V}=\{ \bgalpha\in \mathbb{R}^{|\blJ| } ; \bm{1}_{|\blJ|}^\top {\bgalpha} =\bm{0} \}$, and define the orthogonal projection matrix
$
\blP_{\mathcal{V} }   =(I-  {| \blJ|^{-1}}\bm{1}_{|\blJ|} \bm{1}^\top_{|\blJ|} ),$
where $I$ denotes the identity matrix of order $|\blJ|$. Notice that $\ell_{\rm Mult}(\bgtheta \blx,\bly )=\ell_{\rm Mult}(\bgtheta' \blx,\bly )$ for any $(\blx,\bly)$ if and only if $\blP_{ \mathcal{V} } \bgtheta =\blP_{ \mathcal{V} } \bgtheta' $.
\subsection{Subspace decomposition}\label{sec:Subspace_Decomposition}
We now introduce the subspace decomposition that allows us to parsimoniously model the mass function of interest. 
Naturally, the dependence between response variables is arbitrarily complex when $\bgtheta \in \mathbb{R}^{|\blJ| \times p}$ without additional constraints. To discover parsimonious association structures, we decompose $\bgtheta$ into a sum of components, each of which spans a particular subspace. Returning to an example from the introduction, when $q = 2$, we can decompose $\bgtheta = \blH_{\{0\}} \bgbeta_{\{0\}} + \blH_{\{1\}} \bgbeta_{\{1\}} + \blH_{\{2\}} \bgbeta_{\{2\}} + \blH_{\{1,2\}} \bgbeta_{\{1,2\}},$ where each $\blH$ is a basis matrix, and $\bgbeta$ are the corresponding coefficients. With appropriately constructed $\blH$, if $\bgbeta_{\{1,2\}} = 0$, then the two response variable are independent. We demonstrate how to construct such bases $\blH$ in the following example.




\begin{example}[Subspace decomposition of a $J_1 \times J_2$ contingency table]\label{example:subspace_decomposition_2}
Consider the intercept only model (i.e., $p = 1$ with $\blx_i = 1$) with \( q = 2 \), and the categorical responses having \( J_1 =2\) categories for the first component and \( J_2 = 3 \) categories for the second. We can write $\bgtheta=(a_{11},\dots,a_{J_1 J_2})^\top \in \mathbb{R}^{J_1 J_2}$ as
\begin{equation}\label{equ:a_E_2}
    \Big(\operatorname{vec}^{-1}_{\blJ} \big( \bgtheta \big)\Big)_{j_1,j_2}=a_{j_1,j_2},
\end{equation}
where $a_{j_1,j_2} \in \mathbb{R}$ for any $(j_1,j_2)\in \blJ$.  
Accordingly, we can rewrite \eqref{equ:a_E_2} as  
\begin{equation*}
    {\bgtheta=\sum_{j_2=1}^{J_2} \sum_{j_1=1}^{J_1}a_{j_1,j_2}\ble_{j_2}^{J_2}\otimes  \ble_{j_1}^{J_1} \text{ and }\operatorname{vec}^{-1}_{\blJ} \big( \bgtheta \big) =  \sum_{j_2=1}^{J_2} \sum_{j_1=1}^{J_1}a_{j_1,j_2} \blE^{\blJ}_{j_1j_2}  }, 
\end{equation*}
where $\ble^{J_i}_{j_i}$ is the $j_i$-th standard basis vector for $\mathbb{R}^{J_i}$ (i.e., the $j_i$-th column of $I_{J_i}$) for $i \in [2],$ {and $\blE^{\blJ}_{j_1j_2}$ denotes the standard basis 2-way array for $\mathbb{R}^{J_1\times J_2}$, which is defined as the array whose \((j_1, j_2)\)-th entry is 1 and all other entries are 0.}


Define $\blU_m$ as a matrix such that $[\frac{1}{\sqrt{m}} \bm{1}_{m},\blU_m ]$ is an orthogonal matrix of order $m$. Let $\mathcal{R}(\blU)$ denote the column space of the matrix $\blU$. Then, we can rewrite $\mathbb{R}^{J_i}$ as the internal direct sum between $\mathcal{R}(\blU_{J_i})$ and $\mathcal{R}(\bm{1}_{J_i})$, denoted as $\mathbb{R}^{J_i}=\mathcal{R}(\blU_{J_i})\oplus \mathcal{R}(\bm{1}_{J_i})$. The definitions of the internal direct sum and tensor product can be found in Section S1 of \cite{zhao2024subspace}.

Due to the bilinearity of tensor product, $  \mathbb{R}^{J_2}\otimes \mathbb{R}^{J_1} $ can be decomposed into an internal direct sum

\begin{align*}
 \mathbb{R}^{J_2} \otimes \mathbb{R}^{J_1} 
    =& \{\mathcal{R}(\blU_{J_2}) \oplus  \mathcal{R}(\bm{1}_{J_2}) \}\otimes\{\mathcal{R}(\blU_{J_1})\oplus \mathcal{R}(\bm{1}_{J_1}) \} \notag \\
     =& \{\mathcal{R} (\bm{1}_{J_2})\otimes  \mathcal{R} (\bm{1}_{J_1})\}
     \oplus \{    \mathcal{R} (\bm{1}_{J_2})\otimes  \mathcal{R} (\blU_{J_1})\} \oplus   \{\mathcal{R} (\blU_{J_2})\otimes  \mathcal{R} (\bm{1}_{J_1}) \} \label{eq:internalDirectsum}\\
     & ~~~~~~~~~~~~~~~~~~~~~~~\hspace{2pt}
     \oplus  \{\mathcal{R} (\blU_{J_2})\otimes  \mathcal{R} (\blU_{J_1})\}\notag.
    \end{align*}

{Consider an isomorphism $T$ from the tensor product $ \mathbb{R}^{J_2}\otimes \mathbb{R}^{J_1} $ to $\mathbb{R}^{|\blJ|}$. The isomorphism $T$ is uniquely determined by the change of basis $T(\otimes_t( \ble_{j_2}^{J_2},\ble_{j_1}^{J_1}))=
 \ble_{j_2}^{J_2}\otimes  \ble_{j_1}^{J_1}$, where the tensor product $\otimes_t(\blu, \blv)$ denotes the bilinear map of $(\blu, \blv)$ from the Cartesian product $ \mathbb{R}^{J_2}\times \mathbb{R}^{J_1}$, whose basis can be chosen as $\{ \otimes_t( \ble_{j_2}^{J_2},\ble_{j_1}^{J_1});  1\leq j_1\leq J_1,1\leq j_2\leq J_2\}.$}

Applying the isomorphism $T$ onto each subspace, we obtain that 
$ T(   \mathcal{R} (\blV_{2})\otimes  \mathcal{R} (\blV_{1}) ) = \mathcal{R}( \blV_2 \otimes \blV_1),
$
where $\blV_i \in \{ \frac{1}{\sqrt{J_i}} \bm{1}_{J_i}, \blU_{J_i}\}$ for $ i \in [2]$.   Hence, for $\bgtheta \in \mathbb{R}^{|\blJ|}$, we can write
\[
\bgtheta= \sum_{\blV_1\in \big\{ \frac{1}{\sqrt{J_1}} \bm{1}_{J_1}, \blU_{J_1} \big\} }\sum_{\blV_2\in \big\{ \frac{1}{\sqrt{J_2}} \bm{1}_{J_2}, \blU_{J_2} \big\} } \Big( \blV_2\otimes \blV_1\Big) \bgalpha_{ \blV_2,\blV_1}
\]
for vectors $\bgalpha_{\blV_2,\blV_1}$ of appropriate size. More simply, we may write 
$$
\bgtheta= \sum_{\blk \in \mathcal{K}} \blH_{\blk }\bgbeta_{\blk}, ~~\text{where}~~ \blH_{\{0\}}=\frac{1}{\sqrt{|\blJ|}}\bm{1}_{|\blJ|} ,~~~ \blH_{\{1\}}=\frac{1}{\sqrt{  J_2}} \bm{1}_{  J_2} \otimes \blU_{J_1},~~~\blH_{\{2\}}=\blU_{J_2} \otimes \frac{1}{\sqrt{  J_1}} \bm{1}_{  J_1},$$$$~~~~~~~~~~~~~~~~~~\blH_{\{1,2\}}=\blU_{J_2} \otimes  \blU_{J_1}, ~~~ \mathcal{K}=\{\{0\},\{1\},\{2\},\{1,2\} \},$$
where each $\bgbeta_{\blk}$ is simply the corresponding $\bgalpha_{\blV_2,\blV_1}$. 
As we will formalize in Lemma \ref{thm:structure}, Lemma \ref{lem:decomposition}, and Theorem \ref{thm:structure}, ${\rm span}(\blH_{\{0\}})$ is the subspace for overall effect, ${\rm span}(\blH_{\{1\}}),{\rm span}(\blH_{\{2\}})$ are the subspaces for marginal effect on $Z_1$ and $Z_2$, respectively, and $
{\rm span}(\blH_{\{1,2\}})$ is the subspace for joint effect on $(Z_1, Z_2)$. Because of this, sparsity in coefficients corresponding to each subspace can imply an interpretable restriction on $\bgtheta.$
\end{example}

The discussion outlined above can be generalized to any $\blJ$, with the corresponding isomorphism, denoted as $T_{\blJ}$, being applicable to each subspace.

\begin{lemma}[Isomorphism]\label{lem:isomorphism_T}
Define an isomorphism $T_{\blJ}$ from the tensor product space $\mathbb{R}^{J_q}\otimes\cdots\otimes \mathbb{R}^{J_1} $ to $\mathbb{R}^{|\blJ|}$, which is uniquely determined by the change of basis \\$T_{\blJ} (\otimes_t(\ble_{j_q}^{J_q},\cdots, \ble_{j_1}^{J_1}) )=\ble_{j_q}^{J_q}\otimes \cdots \otimes  \ble_{j_1}^{J_1}$. Here, $\{ \otimes_t(\ble_{j_q}^{J_q},\cdots,  \ble_{j_1}^{J_1} )\}_{(j_1,\cdots,j_q) \in \blJ}$ and\\ $\{\ble_{j_q}^{J_q}\otimes \cdots \otimes  \ble_{j_1}^{J_1}\}_{(j_1,\cdots,j_q) \in \blJ}$ denote the basis of $ \mathbb{R}^{J_q}\otimes\cdots\otimes \mathbb{R}^{J_1} $ and $\mathbb{R}^{|\blJ|}$, respectively. Then, 
\begin{enumerate}
    \item[(i)] for any vector $\blv_i\in \mathbb{R}^{J_i},i \in [q]$, we have
\begin{equation}\label{equ:T_vec}
    T_{\blJ}\Big(\otimes_t(\blv_q,\cdots,\blv_1) \Big)=\blv_q\otimes \cdots \otimes \blv_1;
\end{equation}
    \item[(ii)] and for any $\blV_i \in \{ \frac{1}{\sqrt{J_i}} \bm{1}_{J_i}, \blU_{J_i}\}$, $i \in [q]$, we have
\begin{equation}\label{equ:T_8_J}
    T_{\blJ}\Big(   \mathcal{R} (\blV_{q})\otimes \cdots \otimes  \mathcal{R} (\blV_{1}) \Big) = \mathcal{R}\Big(   \blV_q \otimes \cdots \otimes \blV_1 \Big).
\end{equation}
\end{enumerate}
\end{lemma}

\noindent In Lemma~\ref{lem:isomorphism_T}, the isomorphism \(T_{\blJ}\) is not only between vector spaces \( \mathbb{R}^{J_q} \otimes \cdots \otimes \mathbb{R}^{J_1} \) and \(\mathbb{R}^{|\blJ|}\), i.e., 
\begin{equation*}
    T_{\blJ}(\blv + \blw) = T_{\blJ}(\blv) + T_{\blJ}(\blw), \quad T_{\blJ}(a \blv) = a \cdot T_{\blJ}(\blv), \quad \blv, \blw \in \mathbb{R}^{J_q} \otimes \cdots \otimes \mathbb{R}^{J_1}, \quad a \in \mathbb{R},
\end{equation*}
but also preserves the bilinear function, i.e., the statement of \eqref{equ:T_vec} holds. Here, the Kronecker product serves as the bilinear function.


To summarize Example~\ref{example:subspace_decomposition_2} and Lemma~\ref{lem:isomorphism_T}, we define $\blU_m$ as any matrix such that $[\frac{1}{\sqrt{m}}\bm{1}_m ,\blU_m]$ is an orthogonal matrix of order $m$. Without loss of generality, we take
\begin{equation*}
\blU_m= \left[ \frac{(1,-1,0,\cdots,0)^\top}{\sqrt{ 2}} ,\frac{(1,1,-2,0,\cdots,0)^\top}{\sqrt{6}}  ,\cdots,\frac{(1,\cdots,1,-(m-1))^\top}{\sqrt{(m-1) m}}  \right].   
\end{equation*}
Define the index space $\mathcal{K}_s=\{\blk=\{k_1,\cdots,k_s\} \subset [q] ;  1\leq k_1<k_2<\cdots <k_s\leq q \}$ for $s \in [q]$.  Define the order of the \( \blk \)-interaction as \(\|\blk\|_0=s\) and its number of parameters as \(|\blk|_{J} = (J_{k_1} - 1) \cdots (J_{k_s} - 1)\), if \( \blk \in \mathcal{K}_s\). Thus, the space $\mathcal{K}_s$ defines the set of all possible joint effects of order $s$. Let $\mathcal{K}_0=\{\{0\}\}$, and define $\norm{\blk}_0=0$ and $|\blk|_{\blJ}=1$ when $\blk={0}$. In this context, $\blk=\{0\}$ is treated as an empty set. 
The number of parameters (per predictor) for all $s$-th order effects is given by $L_s=\sum_{1\leq i_1<\cdots<i_s\leq q} (J_{i_1}-1)\cdots (J_{i_s}-1), s \in [q]$. Let $L_0=1$. Define
\[
\blH_0=\frac{\bm{1}_{ |\blJ| }}{ \sqrt{ |\blJ| } } =\frac{\bm{1}_{J_{q} }}{ \sqrt{J_{q} } } \otimes\frac{\bm{1}_{J_{q-1} }}{ \sqrt{J_{q-1} } } \otimes  \cdots \otimes \frac{\bm{1}_{J_{2} }}{ \sqrt{J_{2} } } \otimes \frac{\bm{1}_{J_{1} }}{ \sqrt{J_{1} } }.
\]
For any $\blk=\{k_1,\cdots,k_s\}\in \mathcal{K}_s,s\geq 1$, define 
\begin{equation}\label{def:H_k}
    \blH_{ \blk }= \blV_q \otimes  \blV_{q-1}\otimes  \cdots  \otimes \blV_{2} \otimes \blV_{1}, \qquad \blV_i = \left\{ \begin{array}{ll} 
    \blU_{J_i} & i\in \blk\\
    \frac{\bm{1}_{J_{i} }}{ \sqrt{J_{i} } } & i\in [q]\backslash \blk
    \end{array}\right. .
\end{equation}

Following from Example \ref{example:subspace_decomposition_2}, we see that ${\rm span}(\blH_{\blk})$ is the subspace corresponding to the $\blk=\{k_1,k_2,\cdots,k_s\}$-joint effects. Importantly, it can be verified that the columns of $\blH_{\blk}$ are orthonormal. 


\begin{lemma}[Subspace decomposition]\label{lem:decomposition}
We can express $\mathbb{R}^{|\blJ|}$ as the orthogonal direct sum of the family $\{\mathcal{R}(\blH_{\blk})\}_{\blk\in \cup_{s=0}^q \mathcal{K}_s }$ of subspaces of $\mathbb{R}^{|\blJ|}$, where $\mathcal{R}(\blH_{\blk})$ denotes the column space of $\blH_{\blk}$ and the orthogonal direct sum is defined in Section~S1 of \cite{zhao2024subspace}. 
Furthermore, for any $\blk\in \cup_{s=0}^q \mathcal{K}_s$, the orthogonal projection matrix onto $\mathcal{R}(\blH_{\blk})$ is given by $\blH_{\blk}\blH^\top_{\blk}$.     
\end{lemma}
\noindent Lemma \ref{lem:decomposition} has two key implications. First, it ensures invariance of our proposed estimator {(Subsection \ref{sebsec:inv})}, and second, it allows us to identify the effect encoded by the span of each $\blH_{\blk}$, as we discuss in the following subsection.

\subsection{Reparameterization via subspace decomposition}\label{sec:Reparameterization}
When fitting \eqref{def:multinomial_prob}, it is common to restrict the hypothesis space of models to include joint effects of at most order $d$ for some $0 \leq d \leq q.$ For example, if $q = 3$, we only want to consider models with all possible main effects and two-way interaction effects. In this case, we would take $d = 2$. As such, define the space of possible effects is $\mathcal{K} =\cup_{s=0}^d\mathcal{K}_s$. We call $\mathcal{K}$ the association index space. Though $\mathcal{K}$ is a function of $d$, the maximal order of effect considered, we omit notion indicating this dependence for improved display. 

The following theorem elucidates how the reparameterization of $\bgtheta$ through our subspace decomposition neatly characterizes relationships of mutual, joint, and conditional independence among categorical responses.
  
\begin{theorem}[Sparsity and interpretable models]\label{thm:structure}
Let $\mathcal{I}_1,\mathcal{I}_2,\cdots,\mathcal{I}_m$ be a partition of $[q]$. For any $\blk \in \mathcal{K}_s$, let $\bgbeta_{\blk}$ be a matrix in $\mathbb{R}^{|\blk|_{\blJ} \times p}$. For any $\mathcal{I}=\{ i_1,\cdots,i_s\}\subset [q]$ such that $1\leq i_1< \cdots< i_s\leq q$, define $\blJ^{\mathcal{I}}=  [J_{i_1}] \times \cdots \times [J_{i_s}]$. 
\begin{enumerate}
    \item (Mutual and joint independence) Let \(\mathcal{S}_{\rm joint}= \{\blk \in \mathcal{K} : \exists \ i\in [m] \text{ such that } \blk\subset \mathcal{I}_i  \}.\)
If $ \bgbeta_{\blk}=\bm{0}$ for all $\blk \not\in \mathcal{S}_{\rm joint}$, i.e., the parameter $\bgtheta$ under either Poisson \eqref{def:poisson_mean} categorical response model or multinomial \eqref{def:multinomial_prob} categorical response model is given by $\bgtheta = \sum_{\blk\in \mathcal{K}} \blH_{\blk} \bgbeta_{\blk} = \sum_{\blk\in \mathcal{S}_{\rm joint}} \blH_{\blk} \bgbeta_{\blk},$  then 
\begin{equation}\label{equ:joint_indp}
    \pi_{\blj}(\blx)=\prod_{l=1}^m\pi_{\blj_{\mathcal{I}_l} ,+}(\blx), 
\end{equation}
where $\blj_{\mathcal{I}_l}=\{j_{i^l_1},  \cdots,j_{i^l_s}\}\in \blJ^{\mathcal{I}^l}$, $\mathcal{I}^l=\{ i^l_1, \cdots,  i^l_s \} $ for some $s$ (which may depend on $l$), and $\pi_{\blj_{\mathcal{I}_l} ,+}$ is the marginal pmf of the responses corresponding to $\mathcal{I}_l$.
    \item (Conditional independence) Let \( \mathcal{S}_{\rm joint|\mathcal{I}_m}= \{\blk \in \mathcal{K} : \exists \ i \in [m-1] \text{ such that } \blk\subset \mathcal{I}_i\cup\mathcal{I}_m\}.\) If $ \bgbeta_{\blk}=\bm{0}$ for all $\blk \not\in \mathcal{S}_{\rm joint|\mathcal{I}_m }$, i.e., the parameter $\bgtheta$ under either Poisson \eqref{def:poisson_mean} categorical response model or multinomial \eqref{def:multinomial_prob} categorical response model is given by $\bgtheta = \sum_{\blk\in \mathcal{K}} \blH_{\blk} \bgbeta_{\blk} = \sum_{\blk\in \mathcal{S}_{\rm joint|\mathcal{I}_m  } } \blH_{\blk} \bgbeta_{\blk},$  then 
\begin{equation}\label{equ:cond_indp}
    \pi_{\blj_{\mathcal{I}_1 },\blj_{\mathcal{I}_2 },\cdots,\blj_{\mathcal{I}_{m-1} }|\blj_{\mathcal{I}_m }   }(\blx)=\prod_{l=1}^{m-1}\pi_{\blj_{\mathcal{I}_l} ,+|\blj_{\mathcal{I}_m }}(\blx),
\end{equation}
where
\[
\pi_{\blj_{\mathcal{I}_1 }, \cdots,\blj_{\mathcal{I}_{m-1} }|\blj_{\mathcal{I}_m }   }(\blx)=\frac{\pi_{  \blj    }(\blx)}{\pi_{ \blj_{\mathcal{I}_m }, +   }(\blx)},~\text{ and }~\pi_{\blj_{\mathcal{I}_l }+  |\blj_{\mathcal{I}_m }   }(\blx)=\frac{\pi_{  \blj_{\mathcal{I}_l \cup \mathcal{I}_m },+    }(\blx)}{\pi_{ \blj_{\mathcal{I}_m }, +   }(\blx)}.
\]

\end{enumerate}    
\end{theorem}

\noindent To illustrate the practical implications and applications of Theorem~\ref{thm:structure}, we present the following example. This example is specifically designed to clarify the theorem's underlying principles and to showcase its utility within a hierarchical model, {see Section~\ref{subsec:hierarchical_model}}.

\begin{example}\label{example:3scheme}
Suppose $q=4$. The following types of dependence structures---akin to those in Chapter 6 of \cite{mccullagh1989generalized}---are encoded in the sparsity of the $\bgbeta_{\blk}$. Recall that the random multivariate categorical response is $(Z_1,\cdots,Z_q)\in \blJ$.
\begin{enumerate}
    \item \textbf{Mutual independence.}
        If $\bgtheta= \blH_{\{0\} }\bgbeta_{ \{0\} } +\blH_{\{ 1\}}\bgbeta_{ \{1\} } +\blH_{\{2\}}\bgbeta_{ \{2\} } +\blH_{ \{3\} }\bgbeta_{ \{3\} }+\blH_{ \{4\} }\bgbeta_{ \{4\} },$ then $Z_1,Z_2,Z_3$ and $Z_4$ are mutually independent for any given $\blx$, i.e., for all $\blx \in \mathcal{X}$
    \begin{equation*}
        \pi_{j_1,j_2,j_3,j_4}(\blx)=\pi_{j_1,+,+,+}(\blx)\cdot \pi_{+,j_2,+,+}(\blx)\cdot \pi_{+,+,j_3+,}(\blx)\cdot \pi_{+,+,+,j_4}(\blx), 
    \end{equation*}
    for all $(j_1,j_2,j_3,j_4)\in \blJ.$
    \item \textbf{Joint independence.} If 
    \begin{equation*}
        \bgtheta= \sum_{i=0}^4\blH_{\{i\}}\bgbeta_{\{i\} }  +\Big(\blH_{\{2,3 \} }\bgbeta_{\{2,3\} }+\blH_{\{2,4\} }\bgbeta_{\{2,4\}}+\blH_{\{3,4\}}\bgbeta_{\{3,4\}}+\blH_{\{2,3,4\}}\bgbeta_{\{2,3,4\}}\Big),
    \end{equation*}
then the variable $Z_1$ is jointly independent of $\{Z_2,Z_3,Z_4\}$ for any given $\blx$, i.e., for all $\blx \in \mathcal{X}$
    \begin{equation*}
        \pi_{j_1,j_2,j_3,j_4}(\blx)=\pi_{j_1,+,+,+}(\blx)\cdot \pi_{+,j_2,j_3,j_4}(\blx), \text{ for all } (j_1,j_2,j_3,j_4)\in \blJ.
    \end{equation*}
    \item \textbf{Conditional independence.}  If 
    \begin{equation*}
        \bgtheta= \sum_{i=0}^4\blH_{\{i\}}\bgbeta_{\{i\} }  +\Big(\sum_{2\leq i<j\leq 4}\blH_{\{i,j\}}\bgbeta_{\{i,j\}}+ \blH_{\{2,3,4\}}\bgbeta_{\{2,3,4\}} \Big)+\blH_{\{1,4\}}\bgbeta_{\{1,4\}}, 
    \end{equation*}
    then the variable $Z_1$ and $\{Z_2,Z_3\}$ are conditionally independent for any given $\blx$ and $Z_4$, i.e., for all $\blx \in \mathcal{X}$,
    \begin{equation*}     
\pi_{j_1,j_2,j_3|j_4}(\blx)= {\pi_{j_1,+,+|j_4}(\blx)\cdot \pi_{+,j_2,j_3|j_4}}(\blx),  \text{ for all } (j_1,j_2,j_3,j_4)\in \blJ.
    \end{equation*}
\end{enumerate}    
\end{example}

\noindent The neat interpretations in Example 2 rely partly on a hierarchical structure of the effects. That is, high-order effects are included only if all the corresponding low-order effects are included. Formally, if effect $\blk$ is included in the model, then all $\blk' \in \mathcal{K}$ such that $\blk' \subset \blk$ must also be included in the model. For example, with $q = 3$, if the joint effect $\{1, 2, 3\}$ is included in the model, then for the hierarchy to be enforced, the effects $\{0\}, \{1\}, \{2\},  \{3\}, \{1,2\}, \{1,3\},$ and $\{2,3\}$ must all be included in the model. 

{Formally, given an association index space $\mathcal{K}$, the corresponding class of \textit{hierarchical association index space} is the collection of all sets $\mathcal{N} \subset \mathcal{K}$ such that if
$\blk \in \mathcal{N}$, then $\mathcal{P}(\blk) \subset \mathcal{N},$
where $\mathcal{P}(\blk)$ denotes the powerset of $\blk$ (with the null set replaced with $\{0\}$). }

To restrict attention only to models that respect such a hierarchy, it is natural to consider a class of hierarchical hypotheses spaces
\begin{equation}\label{equ:Hierarchical_hypotheses_space}
   \left\{\bgtheta \in \mathbb{R}^{|\blJ| \times p}: \mathcal{R}(\bgtheta) =\sum_{\blk \in  \mathcal{N}} \mathcal{R}(\blH_{\blk}) \subset \mathbb{R}^{|\blJ|}, ~~\mathcal{N}\subset \mathcal{K} ~\text{ s.t. }~ {\blk\in \mathcal{N} \implies \mathcal{P}(\blk) \subset \mathcal{N}} \right\}. 
\end{equation}


In the next subsection, we will propose a penalized maximum likelihood estimator that allows to explore models in $\mathcal{K}$ or its corresponding hierarchical association index space. 


\section{Penalized likelihood-based association learning}\label{sec:pen_L_E}
\subsection{Penalized maximum likelihood estimation}

Define the negative log-likelihoods as $L^{{\rm Mult}}_n$, and its reparametarized versions $\mathcal{L}^{\rm Mult}_n$  where $L_n^{\rm Mult}(\bgtheta) = \frac{1}{n}\sum_{i=1}^n \ell_{\rm Mult} (\bgtheta \blx_i,\bly_i)$ and {$\mathcal{L}^{\rm Mult}_n(\bgbeta) = L^{\rm Mult}_n(\blH \bgbeta)$. Similarly define $\mathcal{L}^{\rm Pois}_n(\bgbeta) = L_n^{\rm Pois}(\blH \bgbeta)$.}
To simplify the notation and unify the statements and analysis, set
\begin{equation*}
    \mathcal{L}_n( \bgbeta ) = \left\{ \begin{array}{ll} 
    \mathcal{L}^{\rm Mult}_n(\bgbeta) & :\ell=\ell_{\rm Mult} \\
    \mathcal{L}^{\rm Pois}_n(\bgbeta) & :\ell=\ell_{\rm Pois}
    \end{array}\right. .
\end{equation*}
As described in the previous section, due to our subspace decomposition, association structure learning is achieved by learning the sparsity pattern of $\bgbeta\in \mathbb{R}^{ \sum_{s=0}^dL_s \times p }$. For this, we will use penalized maximum likelihood estimators of the form
\begin{equation}\label{def:estimator}
    \widehat\bgbeta\in \argmin_{\bgbeta    } \left\{ \mathcal{L}_n(\bgbeta)+ \lambda \Omega(\bgbeta)\right\},
\end{equation}
for convex penalties $\Omega:\mathbb{R}^{ \sum_{s=0}^dL_s \times p } \to [0,\infty)$ to be discussed in the next subsection. 

\subsection{Global versus local association learning}
Given {$d\in [q]$ and association index space $\mathcal{K}$ (determined by $d$),} to take the advantages of the subspace decomposition in Section~\ref{sec:2subspace_learning}, we parameterize $\bgtheta$ as 
\begin{equation}\label{def:theta_and_H}
    \bgtheta= \sum_{\blk \in \mathcal{K}} \blH_{\blk}\bgbeta_{\blk} =: \blH \bgbeta, 
\end{equation}
$\text{with }\blH=\{\blH_{\blk}\}_{\blk\in \mathcal{K}}\in \mathbb{R}^{|\blJ|\times \sum_{s=0}^d L_s }\text{ and }\bgbeta=\{\bgbeta_{\blk}\}_{\blk\in \mathcal{K}}\in\mathbb{R}^{ \sum_{s=0}^d L_s \times p}.$ Let $\blx_{i(j)} \in \mathbb{R}^{p_j}$ be the $j$th subvector of $\blx_i$, $j \in [t]$, where $\sum_{j=1}^t p_j=p$. Without loss of generality, we partition the matrix $\bgbeta_{\blk}=[\bgbeta_{\blk,1},\bgbeta_{\blk,2},\cdots,\bgbeta_{\blk,t}]$ and vector $\blx_i = (\blx_{i(1)}^\top, \dots, \blx_{i(t)}^\top)^\top \in \mathbb{R}^{p}$ so that 
\begin{equation*}
\bgtheta \blx_i =  \sum_{\blk \in \mathcal{K}} \sum_{j=1}^t\blH_{\blk} \bgbeta_{ {\blk},j} \blx_{i(j)},~~~~~~~\bgbeta_{ \blk,j}\in \mathbb{R}^{ |\blk|_{\blJ}\times p_j}.
\end{equation*}
As discussed in the previous section, if $\bgbeta_{\blk} = 0$, then the corresponding effect defined by $\blk$ is not included in our model. Our predictor grouping structure allows us to perform association learning at distinct resolutions: \textit{global association learning} or \textit{local association learning} (i.e., predictor-wise association learning). 

The goal of global association learning is to discover effects such that all predictors contribute to the effect, or none contribute to the effect. For global association learning, we take $t = 1$.  To encourage sparsity in our fitted model so as to discover a small number of global associations, we use a group lasso-type penalty \citep{yuan2006model} with a positive set of (user-specified) weights $\{w_{\blk}\}_{\blk\in \mathcal{K}}$ for $\bgbeta$ and $\bgtheta$, respectively, as
\begin{equation}\label{def:beta_GL}
    \Omega_{\rm global}(\bgbeta)=\sum_{\blk\in \mathcal{K}}  w_{\blk} \norm{\bgbeta_{\blk}},
\end{equation}
\begin{equation}\label{equ:Phi_def}
    \Phi_{\rm global}( \bgtheta ) = \inf_{\bgtheta=\blH \bgbeta} \Omega_{\rm global}(\bgbeta)  = \Omega_{\rm global}(\blH^\top \bgtheta).
\end{equation}
Given that $\bgtheta=\blH \bgbeta$ 
uniquely determines a $\bgbeta\in\mathbb{R}^{ \sum_{s=0}^dL_s \times p }$, the infimum in \eqref{equ:Phi_def} can be omitted.
  Because the Frobenius norm is nondifferentiable at the matrix of zeros, using $\Omega_{\rm global}$ as a penalty can encourage estimates of the $\bgbeta$, $\widehat\bgbeta$ such that $\widehat{\bgbeta}_{\blk} = 0$ for many $\blk \in \mathcal{K}.$

In local association learning, we relax the assumption that all predictors either contribute to an effect, or no predictors contribute to an effect. For example, when $q = 2$, it is possible that for the majority of predictors (but not all), a change in the predictor's value does not lead to a change in any of the local odd-ratios between response variables (i.e., these predictors only affect the marginal distributions of the response).  This was exactly the type of association learning performed by \citep{molstad2023likelihood}. Our local association learning is much more general: we can discover which predictors modify certain high-order effects, and which predictors (or groups of predictors) only affect lower-order effects. 

To achieve this type of learning, define the set $\mathcal{G}_{\rm local} =\{ (\blk, j): \blk\in \mathcal{K} , j \in [p]\},$ let $\{w_{\blk,j}\}_{({\blk,j}) \in \mathcal{G}_{\rm local}}$ be a positive sequence, and define the penalty function
\begin{equation}\label{def:beta_local}
    \Omega_{{\rm local}}(\bgbeta)=\sum_{(\blk, j) \in \mathcal{G}_{\rm local}}  w_{\blk,j} \norm{\bgbeta_{\blk,j}},
\end{equation}
and similarly for $\Phi_{{\rm local}}$. In contrast to $\Omega_{{\rm global}}$, $\Omega_{{\rm local}}$ has nondifferentiabilities when $\bgbeta_{\blk,j} = 0$ for any $(\blk,j) \in \mathcal{G}_{\rm local}$. As such, this penalty can encourage estimates such that $\widehat{\bgbeta}_{\blk,j} = 0$ for many $j \in [p]$, but if $\widehat{\bgbeta}_{\blk,j'} \neq 0$ for any $j'$, then the $\blk$-joint effect is included in the model. 

Defining the set $\mathcal{G} =\{ (\blk, j): \blk\in \mathcal{K} , j \in [t]\}$, and defining $\Omega_{\mathcal{G}}(\bgbeta)=\sum_{(\blk, j) \in \mathcal{G}}  w_{\blk,j} \norm{\bgbeta_{\blk,j}}$, we generalize both global association learning ($t = 1$) and predictor-wide local association learning ($t = p$). More generally, we can perform a version of local association learning with predictors partitioned into $t$ sets. This may be useful, for example, if predictors are categorical and encoded via multiple dummy variables.  

\subsection{Association learning with hierarchical constraints}\label{subsec:hierarchical_model}
As mentioned in Section \ref{sec:Reparameterization}, it is often desirable to enforce a hierarchical structure for the effects. To this end, we can modify both our global and local association structure learning penalties to enforce the hierarchy.  Recall that for the hierarchy to be enforced, we must have that for every effect $\blk$ included in the model, all elements of {$\mathcal{P}(\blk)$} must also be included in the model. 

To achieve model fits of this type, we utilize the overlapping group lasso penalty. 
This penalty is defined by
\begin{equation}\label{def:beta_OGL}
    \Omega^H_{\mathcal{G}} (\bgbeta)=\sum_{(\blk,j)\in \mathcal{G} } w_{\blk,j} \sqrt{ \sum_{\blk':\blk\subset \blk'}\norm{\bgbeta_{  \blk',j}}^2 }
\end{equation} 
and
\begin{equation}\label{def:theta_OGL}
    \Phi_{\mathcal{G}}^H(\bgtheta)=\Omega^H_{\mathcal{G}}(\blH^\top\bgtheta).
\end{equation} 
The term $\sqrt{ \sum_{\blk':\blk\subset \blk'}\norm{\bgbeta_{ {\blk'},j}}^2 }$ is a group lasso penalty on the entire set of coefficients corresponding to effects that include $\blk$ in their powerset. For example, if $q = 3$ and $\blk = \{1\}$, then $\sqrt{\sum_{\blk':\blk\subset \blk'}\|\bgbeta_{\blk',j}\|^2} = \sqrt{\|\bgbeta_{\{1\},j}\|^2 + \norm{\bgbeta_{\{1,2\},j}}^2 + \|\bgbeta_{\{1,3\},j}\|^2 + \|\bgbeta_{\{1,2,3\},j}\|^2}$.  Consequently, this penalty essentially precludes the possibility that $\widehat\bgbeta_{\{1,2\},j} \neq 0$ but $\widehat\bgbeta_{\{1\},j} = 0$, for example, because the penalty enforces $\widehat\bgbeta_{\{1\},j} = 0$ (via nondifferentiability at the origin) only when all higher order effects $\widehat\bgbeta_{\{1,2\},j} = \widehat\bgbeta_{\{1,3\},j} = \widehat\bgbeta_{\{1,2,3\},j} = 0$ as well. See \citep{yan2017hierarchical} for a comprehensive review of how hierarchical structures can be enforced with the overlapping group lasso and related penalties.

\section{Relation to existing work}
\subsection{Alternative parametric links}
Multivariate categorical response regression is a classical problem in categorical data analysis (e.g., see Chapter 6 of \citep{mccullagh1989generalized}). The majority of existing methods designed specifically for this task utilize parametric links between predictors and responses that can yield interpretable fitted models. To best describe these methods, we will first consider the case that $p = 1$ and $\blx_i = 1$ for all $i \in [n]$ (i.e., the analysis of a $q$-way contingency table). 

One popular parametric link is the multivariate logistic transform. This transform maps probabilities $\bgpi \in \mathbb{R}^{|\blJ|}$ to a set of parameters $\bgeta$. These parameters represent the logarithms of the marginal odds, pairwise odds ratios, and higher-order odds ratios, which are derived from all possible joint marginals of subsets $Z_1,\cdots,Z_q$ \citep{mccullagh1989generalized, glonek1996class, molenberghs1999marginal}. For a given $\bgpi$, the transformation $\bgpi\to \bgeta$ can be expressed as a matrix equation:
\begin{equation}\label{equ:log_transform}
\bgeta=C\log (M \bgpi),
\end{equation}
where $C$ is a contrast matrix, and $M$ is a marginalizing matrix that computes the joint marginals from the cell probabilities.  A more general class of log-linear models (where $C$ and $M$ are more general, and $\bgeta = Z \bgeta_z$ for design matrix $Z$), was proposed by \citep{lang1996maximum}.  According to the definition of \cite{bergsma2002marginal}, a numerical value assigned to $\bgeta$ is considered \textit{strongly compatible} if there exists a valid probability distribution $\bgpi$ that corresponds to it. \cite{palmgren1989regression} showed that excluding the cases when $q=2$ with $J_1=J_2$, no explicit solution is available. \cite{glonek1995multivariate} pointed out the difficulty in solving \eqref{equ:log_transform} for the analysis of contingency tables, stating that ``no readily computable criterion, for determining whether a particular $\bgeta$ is valid, is available''. If there are more than two categorical variables, it can happen that no solution exists because of incompatibility of the lower dimensional marginals. 
Evidently, it remains unclear how to determine whether a specific \(\boldsymbol{\eta}\) is strongly compatible. For Bernoulli response $Z_1,\cdots,Z_q$, \cite{qaqish2006multivariate} can determine the strong compatibility of $\bgeta$, and compute $\bgpi$ from a strongly compatible $\bgeta$ using a noniterative algorithm. When any $J_l \geq 3$, however, their results cannot be applied.  

Matters become even more challenging when we consider the more general log-linear regression model $f(\blx_i) =C\log \{M \bgpi(\blx_i)\}$ where $\blx_i \in \mathbb{R}^p$ for linear function $f.$ The goal of our work is to provide an alternative to log-linear models that (i) has parameters that can be interpreted in the same way as log-linear models and (ii) can be easily computed. Desiderata (i) is addressed by Theorem \ref{thm:structure}, and as we will show in a later section, because our estimator is the solution to a convex optimization problem, we can readily employ modern first order methods for (ii). 

\subsection{Generalizing log-linear models for contingency tables}\label{sebsec:inv}
In this section, we will explain how our method generalizes log-linear models used for the analysis of contingency tables. The key is that our method has the interpretability of ``standard'' log-linear models, but our specific subspace decomposition leads to an invariance property that is essential for penalized maximum likelihood-based association learning. 

Log-linear models are a class of statistical models used to describe the relationship between categorical variables by modeling the expected cell counts in a contingency table. These models express the logarithm of expected frequencies as a linear combination of parameters corresponding to main effects and interactions of the variables. Specifically, for a contingency table (i.e., the intercept only model with $p = 1$) with variables \(Z_1\) and \(Z_2\), the model can be written as
\begin{equation}\label{equ:mu_q_2}
    \log(\mu_{j_1j_2}) = \bgLambda_{j_1,j_2} := \mu + \mu_{j_1}^{Z_1} + \mu_{j_2}^{Z_2} + \mu_{j_1j_2}^{Z_1Z_2}
\end{equation}
where \(\mu_{j_1j_2}\) denotes the expected count in cell \((j_1, j_2)\), \(\mu\) is the overall mean, \(\mu_{j_1}^{Z_1}\) and \(\mu_{j_2}^{Z_2}\) represent the main effects of variables \(Z_1\) and \(Z_2\), respectively, and \(\mu_{j_1j_2}^{ Z_1 Z_2 }\) denotes the interaction effect between \(Z_1\) and \(Z_2\).  Under a multinomial sampling scheme, the model can be written as
\begin{equation}\label{equ:pi_q_2}
    \pi_{j_1j_2} =  \frac{\exp (\bgLambda_{j_1,j_2} )}{\sum_{j_1,j_2}\exp (\bgLambda_{j_1,j_2} )}.
\end{equation}
The log-linear model and the multinomial model share the same linear structure of \( \bgLambda_{j_1,j_2} \).


To ensure the parameters in a log-linear model are uniquely estimable, certain constraints must be imposed. Commonly, sum-to-zero constraints are used, where the sum of the main effects and interaction effects for each variable is set to zero. For example, for the main effects, the constraints are:$
\sum_{j_1=1}^{J_1} \mu_{j_1}^{Z_1} = 0$ and $\sum_{j_2=1}^{J_2} \mu_{j_2}^{Z_2} = 0.$
Similarly, for the interaction effects:
\[
\sum_{j_1} \mu_{j_1j_2}^{Z_1Z_2} = 0 \quad \text{for each } j_2 \quad \text{and} \quad \sum_{j_2} \mu_{j_1j_2}^{Z_1Z_2} = 0 \quad \text{for each } j_1.
\]
Alternatively, one could define $\mu_{1}^{Z_1}=0, \mu_{1}^{Z_2}=0$ and $\mu_{j_1j_2}^{Z_1Z_2}=0$ if $j_1=1$ or $j_2=1$. For maximum likelihood estimation (without penalization), the choice of constraint does not matter due to the invariance property of the maximum likelihood estimator. If, on the other hand, one wanted to impose sparsity inducing penalties on the $\mu$, the choice of constraint may affect the solution.


To see this, recall that \(\bgmu^{\blJ}=\{\mu_{j_1,j_2} \}_{ (j_1,j_2) \in \blJ}\) for log-linear model. Let $\blU'_m=[\ble^{m}_2,\cdots,\ble^{m}_m]$.  Similar to $\blH_{\blk}$ defined in \eqref{def:H_k}, for any $\blk=\{k_1,\cdots,k_s\}\in \mathcal{K}_s,s\geq 1$, define 
\begin{equation}\label{def:H_k_prime}
    \blH'_{ \blk }= \blV_q \otimes  \blV_{q-1}\otimes  \cdots  \otimes \blV_{2} \otimes \blV_{1}, \qquad \blV_i = \left\{ \begin{array}{ll} 
    \blU'_{J_i} & i\in \blk\\
    {\bm{1}_{J_{i} }}  & i\in [q]\backslash \blk
    \end{array}\right. .
\end{equation} 
We can thus rewrite \eqref{equ:mu_q_2} in matrix form as
\begin{equation*}
    \operatorname{vec}_{\blJ}\{ \log (\bgmu^{\blJ}) \} = \blH'_{\{0\}}\bgbeta'_{\{0\}}+\blH'_{\{1\}}\bgbeta'_{\{1\}}+\blH'_{\{2\}}\bgbeta'_{\{2\}}+\blH'_{\{1,2\}}\bgbeta'_{\{1,2\}},
\end{equation*}
where \(\{\blH'_{\blk}\}_{\blk}\) are defined in \eqref{def:H_k_prime} with \(q=2\). 
Similarly, recall that \(\bgpi^{\blJ}=\{\pi_{j_1,j_2} \}_{ (j_1,j_2) \in \blJ}\) for the {multinomial log-linear model} so that
\begin{equation*}
    \operatorname{vec}_{\blJ}\{  \bgpi^{\blJ} \} = \frac{\exp(\bgtheta)}{ \innerpoduct{\exp(\bgtheta)}{\bm{1}_{ J_1J_2 }} }, ~~~~\bgtheta=\blH'_{\{0\}}\bgbeta'_{\{0\}}+\blH'_{\{1\}}\bgbeta'_{\{1\}}+\blH'_{\{2\}}\bgbeta'_{\{2\}}+\blH'_{\{1,2\}}\bgbeta'_{\{1,2\}}.
\end{equation*}
Here, \(\blH'_{\{0\}}\bgbeta'_{\{0\}}, \blH'_{\{1\}}\bgbeta'_{\{1\}}, \blH'_{\{2\}}\bgbeta'_{\{2\}}, \blH'_{\{1,2\}}\bgbeta'_{\{1,2\}}\) are the matrix forms of \(\mu\), \(\mu_{j_1}^{Z_1}\), \(\mu_{j_2}^{Z_2}\), and \(\mu_{j_1 j_2}^{Z_1 Z_2}\), respectively for both log-linear model and multinomial model. 
Evidently the log-linear model can be parameterized as $\bgtheta= \sum_{\blk \in \mathcal{K}} \blH'_{\blk}\bgbeta'_{\blk}$.  If we wanted to impose sparsity on the $\bgbeta'$, it would be tempting to use the same group lasso penalty as defined before, 


However, when considering $\Phi'(\bgtheta)=\sum_{\blk} \norm{\bgbeta'_{\blk}}$, we see that $\Phi'(\cdot)$ is not invariant under the choice of identifability constraints. To be more specific, if $\max_{i\in [q]}{J_i}>2$, $\blU''_m=[\ble^{m}_1,\cdots,\ble^{m}_{m-1}]$, and we define $\blH''$ accordingly, then
\begin{equation*}
    \blH'_{\blk}\bgbeta'_{\blk}\not \equiv   {\blH''_{\blk}\bgbeta''_{\blk}} ,\sum_{\blk} \norm{\bgbeta'_{\blk}}\not \equiv \sum_{\blk} \norm{\bgbeta''_{\blk}}, \text{where } \bgtheta= \sum_{\blk \in \mathcal{K}} \blH'_{\blk}\bgbeta'_{\blk}= \sum_{\blk \in \mathcal{K}} \blH''_{\blk}\bgbeta''_{\blk}.
\end{equation*}
{Choosing $\blH''$ instead of $\blH'$ changes how the $\blk$-joint effect influences the categorical response, leading to results that may depend on this arbitrary selection rather than reflecting an inherent property.  }

{To address the invariance issue, one might consider using an overparameterized version of the log-linear model with penalization of the parameters. However, this leads to an explosion in the number of parameters, and the parameter are more difficult to interpret. Moreover, statistical analysis of such an estimator is fundamentally more difficult than the analysis of our estimator.}

In our reparameterization $\bgtheta= \sum_{\blk \in \mathcal{K}} \blH_{\blk}\bgbeta_{\blk}$, the corresponding group lasso penalty $\Phi(\bgtheta)=\sum_{\blk} \norm{\bgbeta_{\blk}}$ is invariant under different choice of $\blU_m$ such that $[\frac{1}{\sqrt{m}}\bm{1}_m ,\blU_m]$ is a real orthogonal matrix. To be more specific, if we let $\mathcal{U}_m$ be another real matrix such that $[\frac{1}{\sqrt{m}}\bm{1}_m ,\mathcal{U}_m]$ is a real orthogonal matrix, and define $\blH_{\blk}^{\mathcal{U}}$ by replacing $\blU_{J_i}$ with $\mathcal{U}_{J_i}$ in \eqref{def:H_k}, then
\begin{equation*}
      \blH_{\blk}\bgbeta_{\blk}\equiv  \blH^{\mathcal{U}}_{\blk}\bgbeta^{\mathcal{U}}_{\blk},\sum_{\blk} \norm{\bgbeta_{\blk}} \equiv \sum_{\blk} \norm{\bgbeta^{\mathcal{U}}_{\blk}}, \text{where } \bgtheta= \sum_{\blk \in \mathcal{K}} \blH_{\blk}\bgbeta_{\blk}= \sum_{\blk \in \mathcal{K}} \blH^{\mathcal{U}}_{\blk}\bgbeta^{\mathcal{U}}_{\blk}.
\end{equation*}

\subsection{Modern approaches to multivariate categorical response regression in high dimensions}
Existing methods for multivariate categorical response regression with a large number of predictors, responses, and/or a large number categories per response typically rely on latent variable models \citep[e.g., the regularized latent class model of ][]{molstad2022conditional}, or classifier chains \citep{read2021classifier}. 

The latent class model is able to capture complex relationships between responses by assuming that given a latent variable $W$, $Z_m$  and $Z_{m'}$ are independent given $\mbX$, i.e., $Z_m \indep Z_{m'} \mid \mbX, \mbW$. Thus, fitted model coefficients cannot be straightforwardly interpreted in terms of the distribution of interest $Z_1, \dots, Z_q \mid \mbX$, as can the coefficients from our fitted model. Moreover, the order of effects in the latent class method cannot, generally speaking, be easily identified unless the effect is null. 

Along similar lines, it is common to decompose the joint mass function of interest into simpler, estimable parts. Methods utilizing to this approach include those most popular in the machine learning literature on ``multilabel classification'' \citep{herrera2016multilabel}, namely, classifier chains \citep{read2021classifier}. A classifier chain estimates $Z_1, \dots, Z_q \mid \mbX$ by fitting a model for $Z_1 \mid \mbX$, then $Z_2 \mid \mbX, Z_1$, then $Z_3 \mid \mbX, Z_1, Z_2$, and so on, and using their product as an estimate of the mass function of interest. This approach requires many ad-hoc decisions that can have a significant impact on how the model performs (e.g., in what order to fit the chain and how to model each specific conditional distribution). Like the latent class model approach, classifier chains cannot be used to identify the order of effects in a straightforward way, which is the primary motivation for our  work.

\section{Computation}\label{sec:main_computaion}
In this section, we outline a proximal gradient descent algorithm---described in Chapter 4 of \cite{parikh2014proximal}---for computing the group lasso and the overlapping group lasso-penalized estimators.

The proximal gradient descent algorithm can be understood from the perspective of the majorize-minimize principle \citep{lange2016mm}. If there exists some $L>0$ such that for $k$-th iterate $\bgbeta^k$,
\begin{equation}\label{ineq:majorize-minimization}
    \mathcal{L}_n(\bgbeta) + \lambda \Omega(\bgbeta) \leq  \mathcal{L}_n(\bgbeta^k)+\innerpoduct{\nabla \mathcal{L}_n(\bgbeta^k ) }{\bgbeta-\bgbeta^k }+\frac{L}{2  } \norm{\bgbeta-\bgbeta^k }^2 + \lambda \Omega(\bgbeta)
\end{equation}
for all $\bgbeta$,
then, if we define the $(k+1)$th iterate as
\begin{equation}\label{equ:proximal}
    \bgbeta^{k+1}=\argmin_{\bgbeta} \left[ \frac{1}{2} \norm{ \bgbeta-\Big\{\bgbeta^k-\frac{1}{L} \nabla \mathcal{L}_n(\bgbeta^k)  \Big\} }^2+\frac{\lambda}{L} \Omega (\bgbeta)\right],
\end{equation}
we are ensured that the objective function at $\bgbeta^{k+1}$ is no greater than the objective function at $\bgbeta^{k}$ (i.e., the sequence of iterates $\{\bgbeta^k\}_{k=1}^\infty$ have the descent property). 
When $\Omega$ is the group lasso penalty  (i.e., $\Omega_\mathcal{G}$), then the  proximal operator \eqref{equ:proximal} has closed form solution 
\[
{\bgbeta}^{k+1}_g= \max \left(  1- \frac{\lambda w_g }{  \norm{ L\cdot \bgbeta^k_{g}- \frac{\partial }{ \partial \bgbeta_g }\mathcal{L}_n(\bgbeta^k) } } ,0 \right) \Big( \bgbeta^k_{g}- \frac{1}{L}\frac{\partial }{ \partial \bgbeta_g }\mathcal{L}_n(\bgbeta^k) \Big),~~~~g\in \mathcal{G}.
\] 
When $\Omega$ is the overlapping group lasso penalty (i.e., $\Omega^H_\mathcal{G}$), we solve the proximal operator \eqref{equ:proximal} using a blockwise coordinate algorithm \citep{jenatton2011proximal,yan2017hierarchical}.

In Lemma~S6 of \cite{zhao2024subspace}, we show that for all $\bgbeta$ and $\bgbeta'$, $\norm{\nabla \mathcal{L}^{\rm Mult}_n(\bgbeta)-\nabla \mathcal{L}^{\rm Mult}_n(\bgbeta')} \leq (2n)^{-1}\lambda_{\max}(\blX^\top\blX) \norm{\bgbeta-\bgbeta'}$ with $\blX=[\blx_1,\cdots,\blx_n]$, which implies that with $L \geq (2n)^{-1}\lambda_{\max}(\blX^\top\blX)$, \eqref{ineq:majorize-minimization} will hold. 
However, in the case of a Poisson categorical response model, the inequality \eqref{ineq:majorize-minimization} cannot hold globally for any $L$. Therefore, we use a proximal gradient descent algorithm with the step size determined adaptively at each iteration by a backtracking line search. 

More details about tuning parameter selection, as well as the formulation of an accelerated variation of the proximal gradient descent algorithm, can be found in Section~S2 of \cite{zhao2024subspace}. More details about the accelerated proximal gradient descent algorithm can be found in Section~4.3 of \cite{parikh2014proximal} and Algorithm~2 of \cite{tseng2008accelerated}. We present our algorithm and all needed sub-algorithms in Section~S2 of our Supplementary Materials~\cite{zhao2024subspace}.



\section{Statistical Properties}\label{sec:stat_prop}
In this section, we examine the statistical properties of the group lasso estimator, as defined in \eqref{def:estimator}, considering variations in $n$, $p$, and $\blJ$. Let $\bgtheta^*=\blH_{\rm full}\bgbeta^*$ represent the data generation parameter, where $\blH_{\rm full}=\{\blH_{\blk}\}_{\blk\in \cup_{s=0}^q\mathcal{K}_s }\in \mathbb{R}^{|\blJ|\times |\blJ| }$. To establish an error bound, it is necessary to define an identifiable estimand: 
the parameter $\bgbeta^\dagger$. Let the set $\mathcal{F}_{\bgtheta^*}$ denote the set of all $\bgbeta$, which leads to the same probability distribution, that is for multinomial and Poisson categorical response models,
\begin{equation}\label{equ:33_theta}
\begin{split}
    \mathcal{F}_{\bgtheta^*}&=\Big\{\widetilde\bgbeta\in \mathbb{R}^{ \sum_{i=0}^q L_i \times p};  \ell ( \bgtheta^* \blx, \bly )= \ell ( \blH_{\rm full} \widetilde{\bgbeta} \blx, \bly ) , \forall (\blx,\bly)   \Big\}\\
    &=\Big\{\widetilde\bgbeta\in \mathbb{R}^{ \sum_{i=0}^q L_i \times p};  \blP_{\mathcal{V}} \bgtheta^*=\blP_{\mathcal{V}} \blH_{\rm full}\widetilde\bgbeta   \Big\},
\end{split}
\end{equation}
where $\blP_{\mathcal{V}}=I-  {| \blJ|^{-1}}\bm{1}_{|\blJ|} \bm{1}^\top_{|\blJ|}  $ for multinomial categorical response model, and $\blP_{\mathcal{V}}=I$ for Poisson categorical response model. 

Define $ \bgtheta^\dagger=\blP_{\mathcal{V}}\bgtheta^* \text{ and }\bgbeta^\dagger=\blH_{\rm full}^{\top}\bgtheta^\dagger.$ {By Lemma~S8 from \cite{zhao2024subspace},} we know that
\begin{equation}\label{def:true_par}
    \bgbeta^\dagger\in \argmin_{\bgbeta \in \mathcal{F}_{\bgtheta^*} }   \mathcal{L}_n(\bgbeta) +\lambda \Omega_{\mathcal{G}}(\bgbeta)   =\argmin_{\bgbeta \in \mathcal{F}_{\bgtheta^*}  }    \Omega_{\mathcal{G}}(\bgbeta).
\end{equation}
Now, we introduce our assumptions. The first is a standard scaling assumption on the predictors. 
\begin{assumption}[Predictor scaling]\label{ass:1}
    The  predictors are scaled so that for any $1\in[n], j \in [t]$, and $\norm{\blk}_0\leq d$,
      $\norm{\blx_{i(j)}}   \leq w_{\blk,j} C$ for finite constant $C$.
\end{assumption}
\noindent
The following assumption regards the data generating process. 
\begin{assumption}\label{ass:2}
The responses $\bly_i^{\blJ}=\{y^i_{\blj }\}_{\blj \in \blJ}, 1\leq i\leq n$ are independent given $\{\blx_i\}_{i=1}^n$ and generated under (i) the Poisson categorical response model or (ii) the multinomial categorical response model with ($n_i = 1$ for $i \in [n]$, without loss of generality).  
\end{assumption}

\begin{assumption}[Poisson categorical response model]\label{ass:3'}
Under (i), the Poisson categorical response model with $\mathcal{V}=\mathbb{R}^{|\blJ|} $, there exists a finite constant $C_1$ such that 
$\Lambda:=\max_{i \in[n]}\|e^{\bgtheta^\dagger \blx_i}\|_\infty\leq C_1.$
\end{assumption}
\noindent Note that under (ii), the multinomial categorical response model, $\mathcal{V}=\{ \bgtheta\in  \mathbb{R}^{|\blJ | } : \bm{1}_{|\blJ|}^\top \bgtheta=\bm{0} \}$. This is not an assumption, but rather a definition. 

Next, we make an assumption on the curvature of the negative log-likelihood in certain directions: this is commonly known as restricted strong convexity \citep[Definition 9.15 and Theorem 9.36]{wainwright2019high}. Let $\mathcal{E}_n(\Delta \bgtheta):=L_n(\bgtheta^\dagger+\Delta \bgtheta)-L_n(\bgtheta^\dagger ) -\innerpoduct{\nabla L_n(\bgtheta^\dagger )}{\Delta \bgtheta}$. 

\begin{assumption}[Restricted strong convexity]\label{ass:3}
Let $\Phi_{\mathcal{G}}(\bgtheta)=\Omega_{\mathcal{G}}(\blH^\top\bgtheta)$ be the reparameterized group lasso penalty for the association learning. The quantity $\mathcal{E}_n(\Delta \bgtheta)$ satisfies restricted strong convexity $(RSC)$ condition with radius $R>0$, constants $A$ and $C$, and curvature $\kappa>0$, i.e., $\Delta \bgtheta 
\in \mathcal{M}_H:=\{\bgtheta: \bgtheta=\sum_{g\in \mathcal{G}} \blH_{\blk} \bgbeta_{g} \}$,
\begin{equation}\label{condition:RSC}
    \mathcal{E}_n(\Delta \bgtheta) \geq \frac{\kappa}{2} \norm{ \blP_{\mathcal{V}} (  \Delta \bgtheta ) }^2-A\cdot C^2 \Big(  {\frac{\log|\mathcal{G}|}{n} }+ {\frac{\vphantom{1}m}{n}}   \Big) \cdot \inf_{\blP_{\mathcal{V}}\Delta \bgtheta = \blP_{\mathcal{V}}\Delta \bgtheta' }\Phi_{\mathcal{G}}^2(\Delta \bgtheta'), \norm{\Delta \bgtheta}\leq R,
\end{equation}
where $|\mathcal{G}|$ is the cardinality of $\mathcal{G}$, and $m=\max_{(\blk,j)\in \mathcal{G}}  |\blk|_{\blJ}\cdot p_j$. Under (i), the Poisson categorical response model,  $\blP_{\mathcal{V}}=I$, and denote $\kappa=\kappa^{Pois}_{\blJ}$, whereas under (ii),  the multinomial categorical response model, $\blP_{\mathcal{V}}=I-{| \blJ|^{-1}}\bm{1}_{|\blJ|} \bm{1}^\top_{|\blJ|}$, and denote $\kappa=\kappa^{Mult}_{\blJ}$. 
\end{assumption}
\noindent The restricted strong convexity condition is a well-understood condition in penalized regression \citep[][Section 2.4]{negahban2012unified}. Effectively, this condition requires that in a neighborhood of the true parameter, the negative log-likelihood has sufficient curvature. 
\begin{remark}\label{rem:strong_convex}
In Lemma~S4 of \cite{zhao2024subspace}, we verify that under mild assumptions on the distribution of predictors, restricted strong convexity holds with high probability for (i) Poisson and (ii) multinomial categorical response models.
\end{remark}


Define the support of $\bgbeta^\dagger$ as $\mathcal{S}=\{ g\in \mathcal{G}; \bgbeta^\dagger_g\neq 0\}$ 
and define $\Psi(\mathcal{S})^2 = \sum_{(\blk,j)\in \mathcal{S}}  {w^2_{\blk,j}}.$ Clearly, 
    if $w_g=1$ for all $g\in \mathcal{G}$, then
$\Psi(\mathcal{S})^2=|\mathcal{S}|.$ Note that $\Psi(\mathcal{S})$ is essentially the subspace compatibility constant \citep[Definition 9.18]{wainwright2019high}: a quantity that often appears in error bounds for regularized M-estimators.  

Note that the dimensionality of $\widehat\bgbeta$ depends on the user-specified $d$, whereas $\bgbeta^\dagger \in \mathbb{R}^{p \times |\blJ|}$. Thus, to simplify notation, let {$\widehat\bgbeta_0$} denote the version of $\widehat\bgbeta$ where all effects of order higher than $d$ have been set to zero (i.e., $\widehat\bgbeta_0 = [\widehat\bgbeta, 0] \in \mathbb{R}^{p \times |\blJ|}$. 
We are now prepared to present our error bound for $\|\widehat\bgtheta-\bgtheta^\dagger\|=\|\widehat\bgbeta_0-\bgbeta^\dagger\|$. Recall that $\mathcal{K}=\cup_{s=0}^d\mathcal{K}_s$, where $d$ denotes the maximal number of association between response variables. Define the true maximal number of association as $d^*=\{ \norm{\blk}_0 ;(\blk,j)\in \mathcal{S} \}$.  
\begin{theorem}\label{thm:main_1}
Let $B, B_1,B_2$ and $B'$ be positive absolute constants, and let $\xi \geq 1$ be fixed.  Suppose that $d$ is chosen so that $d^*\leq d$ and that Assumptions~\ref{ass:1}-\ref{ass:3} hold.  
\begin{enumerate}
    \item[(i)]  Under the Poisson categorical response model, if $\lambda=\xi B C( \sqrt{\Lambda\vphantom{1} m/n}+\sqrt{\Lambda\log|\mathcal{G}|/n})$ with $0\leq(\xi-1)( \sqrt{\vphantom{1} m/n}+\sqrt{\log|\mathcal{G}|/n })\leq B_2$, $\lambda \leq  R \kappa^{Pois}_{\blJ}\{6 \sqrt{|\mathcal{S}|} \}^{-1}$, and $(m/n+\log|\mathcal{G}|/n) \leq B_1 \cdot \min\{1, \kappa^{Pois}_{\blJ}(AC^2 |\mathcal{S}|)^{-1}\}$, then
\[
\|\bgtheta^\dagger- \widehat\bgtheta \|=\|  \bgbeta^\dagger- \widehat\bgbeta_0 \| \leq  \frac{6 \xi B C \sqrt{|\mathcal{S}|} }{ \kappa^{Pois}_{\blJ} } \left( \sqrt{\frac{\vphantom{1} m}{n}}+\sqrt{\frac{\log|\mathcal{G}|}{n} }\right),
\]
with probability at least $1-e^{-B' (\xi-1)^2(m+\log|\mathcal{G}|)}$.    
    \item[(ii)] Under the multinomial categorical response model, if $\lambda=\xi B C ( \sqrt{m/n}+\sqrt{\log|\mathcal{G}|/n}),$  $\lambda \leq R  { \kappa_{\blJ}^{Mult} }  \{6 \sqrt{|\mathcal{S}|} \}^{-1}$, and $(m/n+ \log|\mathcal{G}|/n) \leq B_1 \cdot \min\{1, { \kappa_{\blJ}^{Mult} } (A C^2 |\mathcal{S}|)^{-1}\}$, then
\[
\| \bgtheta^\dagger- \widehat\bgtheta \| = \| \bgbeta^\dagger- \widehat\bgbeta_0 \|\leq  \frac{6 \xi   BC \sqrt{|\mathcal{S}|} }{ \kappa^{Mult}_{\blJ} } \left( \sqrt{\frac{\vphantom{1} m}{n}}+\sqrt{\frac{\log|\mathcal{G}|}{n} }\right),
\]
with probability at least $1-e^{-B' (\xi-1)^2(m+\log|\mathcal{G}|)}$.    
\end{enumerate}
\end{theorem}
\noindent In Lemma~\ref{lem:justify_ass_3} of \cite{zhao2024subspace}, we show that, under certain regularity assumptions, $\kappa_{\blJ}^{Mult} \asymp |\blJ|^{-1}\kappa_{\blJ}^{Pois}$ and $\kappa_{\blJ}^{Pois}=O(1)$.  However, we cannot conclude that the Poisson-likelihood based estimator is better its multinomial counterpart because the data generating models being assumed are fundamentally different.

The result of Theorem \ref{thm:main_1} indicates that under the Poisson or multinomial sampling scheme, assuming $\kappa_{\blJ}^{Pois}=O(1)$ or $\kappa_{\blJ}^{Mult}=(|\blJ|^{-1}) $, we can achieve a Frobenius norm error rate of 
$O(\sqrt{m/n}+\sqrt{\log|\mathcal{G}|/n})$. Call that $|\mathcal{G}|$ is the number of groups of parameters being penalized in \eqref{def:estimator} under general local association learning.  This would seem to suggest that having fewer groups is beneficial, but this term is counterbalanced with $m$, which is the largest number of parameters per group. Hence, since a small number of groups would require a larger number of parameters per group, there is a clear tradeoff between the two. Importantly, both terms are multiplied by $|\mathcal{S}|$, so ideally, we will select a number of groups that leads to small $\mathcal{S}$ without inflating $|\mathcal{G}|$ or $m.$ 

Though not made explicit in our bounds, the effect of a well-specified $d$ is apparent in our error bounds. If $d = q \gg d^*$, then both $m$ and $|\mathcal{G}|$ will be larger than if $d$ were specified closer to $d^*.$ Of course, if $d < d^*$, we could not expect consistent estimation since this will force estimates of truly nonzero effects to be zero.  

The following corollary is a special case of Theorem~\ref{thm:main_1} for multinomial categorical response model, letting $\mathcal{G}=\mathcal{G}_{\rm global}$ or $\mathcal{G}_{\rm local}$. Here, we replace the quantities from Theorem \ref{thm:main_1} with more explicit versions. 
\begin{corol}
Under the conditions of Theorem \ref{thm:main_1}, assuming the multinomial categorical response model, if tuning parameters are chosen in accordance with Theorem \ref{thm:main_1}(ii) and $w_g = 1$ for all $g \in \mathcal{G}$, then
\begin{enumerate}
    \item For global association learning, with probability as specified in Theorem~\ref{thm:main_1}, 
\[
\| \bgtheta^\dagger- \widehat\bgtheta \|  \leq  \frac{6 \xi BC }{ \kappa^{Mult}_{\blJ} }\sqrt{\sum_{\blk \in \mathcal{K}} \mathbf{1}(\bgbeta_{\blk}^\dagger \neq 0)}\left\{\sqrt{\frac{\vphantom{1} p \prod_{\ell=1}^d (J_{(\ell)}-1)}{n}}+\sqrt{\frac{\log\sum_{l=0}^d\binom{q}{l} }{n} }\right\},
\]
where $J_{(1)},\cdots, J_{(q)}$ is a permutation of $J_1,\cdots,J_q$ such that $J_{(1)}\geq J_{(2)} \geq \cdots \geq J_{(q)}$.
    \item For local association learning ($t \geq 2$), with probability as specified in Theorem~\ref{thm:main_1}, 
\[
\|\bgtheta^\dagger- \widehat\bgtheta\| \leq  \frac{6 \xi   BC }{ \kappa^{Mult}_{\blJ} } \norm{\bgbeta^\dagger}_{0,\mathcal{G}}\left\{ \sqrt{\frac{\vphantom{1} \max_{(\blk,j)\in \mathcal{G}}  |\blk|_{\blJ} \cdot p_j}{n}}+\sqrt{\frac{\log  t  + \log \sum_{l=0}^d\binom{q}{l}}{n} }\right\},
\]
where $\norm{\bgbeta^\dagger}_{0,\mathcal{G}}=\sqrt{\sum_{(\blk, j) \in  \mathcal{G}} \mathbf{1}(\bgbeta_{\blk,j}^\dagger \neq 0)}$.

\end{enumerate}\end{corol}
For the multinomial sampling scheme, as $|\blJ|$ increases, the upper bound of the estimation error worsens. This suggests that increasing the dimension of the response will lead to poorer estimation.

We continue by demonstrating the reasonableness of Assumption~\ref{ass:3}, particularly regarding its validity under the assumption of random predictors.
In section 9 of \cite{wainwright2019high}, the restricted strong convexity condition has been derived under a GLM setting (See Theorem 9.36 in \cite{wainwright2019high}). Here, we generalized their results to a multivariate GLM setting, and calibrate the Rademacher complexity term of the group lasso penalty according to multivariate GLM setting.
We summarize the results in S7 of \cite{zhao2024subspace} and incorporate both the multinomial and Poisson categorical response settings into the following lemma.

\begin{lemma}\label{lem:justify_ass_3}
Under Assumptions~\ref{ass:1}--\ref{ass:3'}, assume that $\blx_1, \dots, \blx_n$ are independent and identically distributed with zero mean. Additionally, assume that for some positive constants $(\alpha, \beta)$, we have
\begin{equation*}
    \mathbb{E} \norm{\Delta \bgtheta \blx_i }^2 \geq \alpha \quad \text{and} \quad \mathbb{E} \norm{\Delta \bgtheta \blx_i }^4 \leq \beta,
\end{equation*}
for all vectors $\Delta \bgtheta \in \mathcal{M}_H$ such that $\norm{\Delta \bgtheta} = 1$. Then, the following results hold. For both multinomial and Poisson categorical response models with the reparameterized group lasso penalty $\Phi_{\mathcal{G}}$, the restricted strong convexity condition~\eqref{condition:RSC} in Assumption~\ref{ass:3} holds with probability at least $1-c_1 e^{-c_2 n}$. Furthermore, $\kappa_{\blJ}^{Mult} \asymp |\blJ|^{-1}\kappa_{\blJ}^{Pois}$ and $\kappa_{\blJ}^{Pois}=O(1)$.
\end{lemma}
\noindent The above lemma justifies condition \eqref{condition:RSC} and the typical behavior of the curvature $\kappa$ in Assumption~\ref{ass:3}, showing that both will hold with high probability under mild assumptions.

\section{Numerical studies}\label{sec:simulation} 
\subsection{Data generating models and competitors}
We present a series of simulations designed to evaluate the performance of the proposed methods and applicable variations of existing methods under various scenarios. We consider a range of parameters, including different sample sizes, dimensions, and three different model generation schemes.  A detailed description of this study is in Section S3 of \cite{zhao2024subspace}.\\

\textbf{Parameter setup and simulation.} For $N_{\text{rep}}=100$ independent replications, we simulate data from the multivariate multinomial logistic regression model, with $d=4$, $q=4$, and $(J_1, J_2, J_3, J_4) = (2,2,2,3)$. With $n\in \{100,300,500,1000,2000\}$ training samples, each observation $\blx_i$ is drawn from a multivariate normal distribution ${\rm N}_p(\mathbf{0},\Sigma)$, where the covariance entries $\Sigma_{jk}=0.5^{|j-k|}$ are defined for all pairs $(j,k)\in [p] \times [p]$. Given a coefficient matrix $\bgbeta^* \in  \mathbb{R}^{ \sum_{s=0}^d L_s \times p}$, the probability vector is given by
\[
\operatorname{vec}_{\blJ}\big(\bgpi^{\blJ}_{j_1,j_2,j_3}(\blx)\big) = \frac{\exp(\blH\bgbeta^*\blx)}{\langle \bm{1}_{|\blJ|}, \exp(\blH\bgbeta^*\blx) \rangle},
\]
from which we generate the response  $\bly_i\in \mathbb{R}^p$ as a realization of 
\begin{equation}\label{dist:mult_dist_generate}
    \operatorname{Multinomial}\Big(n_i, \operatorname{vec}_{\blJ}\big(\bgpi^{\blJ}_{j_1,j_2,j_3,j_4}(\blx_i)\big)\Big), \quad n_i=1.
\end{equation}
This process is also extended to generate $1000$ validation samples for model tuning and $N_{\text{test}}=10000$ test samples to evaluate model performance. We conduct our simulations over a range of dimensions $p \in \{10, 50\}$ to assess scalability and robustness. Let $\mathcal{G}_{\rm  global} = \{ (\blk, j); \blk \in \mathcal{K} , j \in \{1,2\} \}$ with $p_1=1,p_2=p-1$ and $\mathcal{G}_{\rm {local}} = \{ (\blk, j); \blk \in \mathcal{K} , j \in \{ 1, \cdots, p \} \}$ with $p_1=p_2=\cdots=p_p=1$.

We consider three distinct structures for $\bgbeta^*$. 
The parameter generation methods for $\bgbeta^*$ are designated as Scheme 1, Scheme 2, and Scheme 3, and are discussed in detail in the Supplementary Materials \cite{zhao2024subspace}. These correspond to the three interpretable models—mutual independence, joint independence, and conditional independence, respectively—as presented in Example~\ref{example:3scheme}. Moreover, in each scheme, the effects are local---only the intercept and two randomly selected predictors have nonzero effects.  Hence, these schemes are most well suited for the versions of our methods using $\Omega_{\rm local}^H$. \\

\textbf{Candidate estimators.} In our simulation studies, we will examine the following eight estimators. The first six estimators—\textrm{O-Mult}, \textrm{O-Pois}, \textrm{L-Mult}, \textrm{L-Pois}, \textrm{G-Mult} and \textrm{G-Pois}—are derived using the reparameterization technique. Here \textrm{O}, \textrm{L}, and \textrm{G} denote estimators using overlapping group lasso with hierarchical structure built on local group $\mathcal{G}_{\rm local}$, group lasso with local group $\mathcal{G}_{\rm local}$, and group lasso with global group $\mathcal{G}_{\rm global}$ penalties, respectively. Additionally, \textrm{Mult} and \textrm{Pois} refer to multinomial and Poisson multivariate categorical response models, respectively. Recall that the data-generating model is based on a multinomial model. Thus, the $\textrm{O-Mult}$, $\textrm{L-Mult}$, and $\textrm{G-Mult}$ are penalized maximum likelihood estimators for a correctly specified model. In contrast, $\textrm{O-Pois}$, $\textrm{L-Pois}$, and $\textrm{G-Pois}$ can be thought of as M-estimators. The seventh estimator, \textrm{G-Mult}-$\bgtheta$, employs the classical parameterization approach in $\bgtheta$. The eighth estimator, \textrm{Sep-Mult}, is designed to individually address each category in the multinomial vector, providing estimates of each response's probability mass function separately. The method denoted \textrm{Oracle} represents the true parameter, and is included to serve as a baseline.\\

\textbf{Tuning criteria.} We employ a train-validation split in order to select tuning parameters in our simulation study. Specifically, we select the candidate tuning parameters that minimize cross-entropy loss on the validation set.

\subsection{Results}
\begin{figure}[tbp]
    \centering
    \begin{minipage}[t]{0.48\linewidth}
    \includegraphics[width=\linewidth]{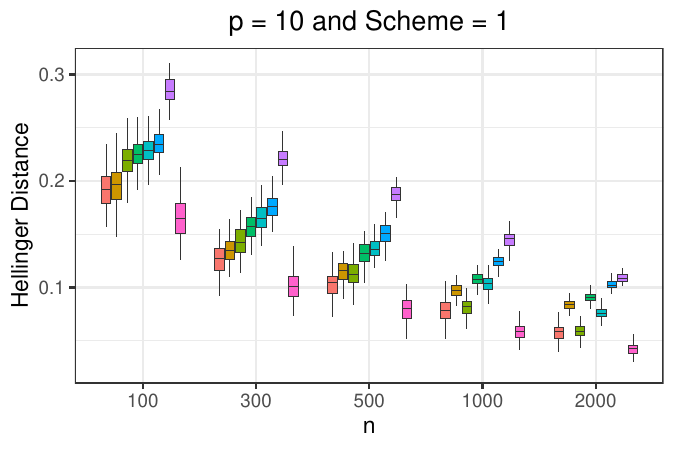}
    \end{minipage}
    \hfill
    \begin{minipage}[t]{0.48\linewidth}
    \includegraphics[width=\linewidth]{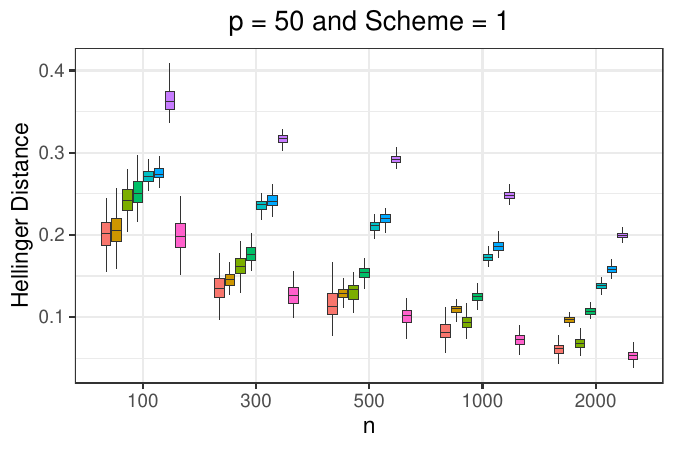}
    \end{minipage}
    
    \begin{minipage}[t]{0.48\linewidth}
    \includegraphics[width=\linewidth]{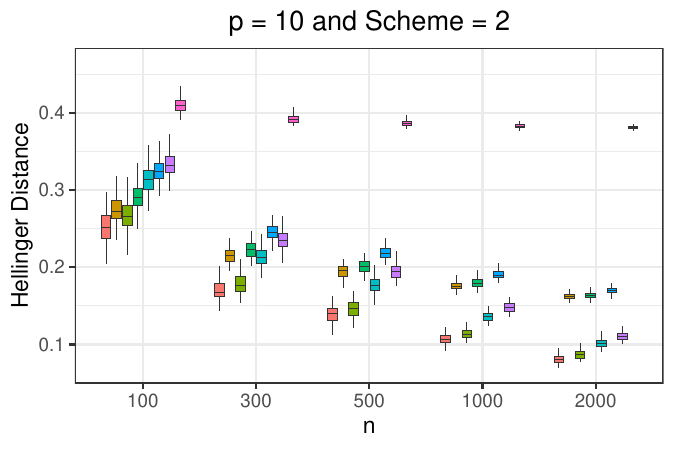}
    \end{minipage}
    \hfill
    \begin{minipage}[t]{0.48\linewidth}
    \includegraphics[width=\linewidth]{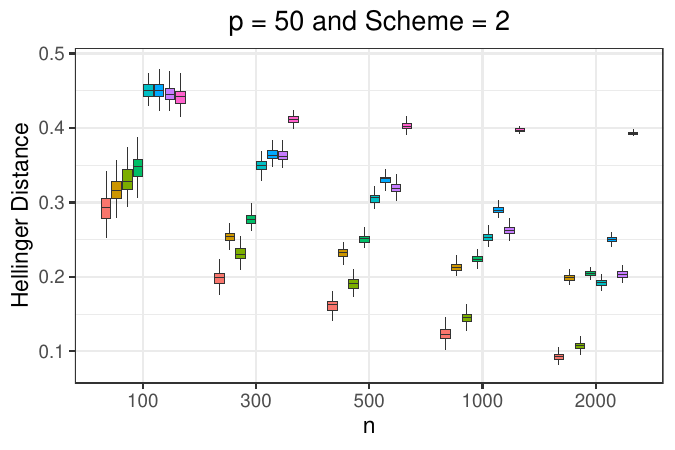}
    \end{minipage}

    \begin{minipage}[t]{0.48\linewidth}
    \includegraphics[width=\linewidth]{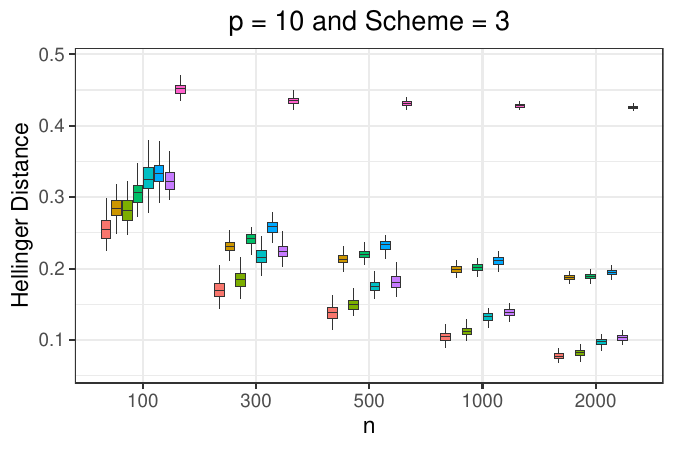}
    \end{minipage}
    \hfill
    \begin{minipage}[t]{0.48\linewidth}
    \includegraphics[width=\linewidth]{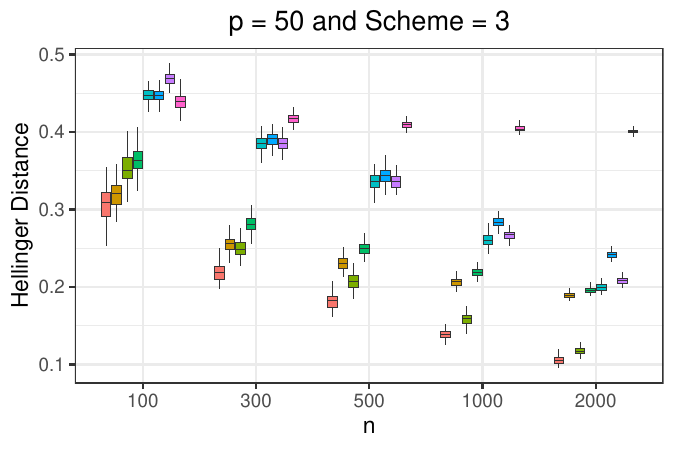}
    \end{minipage}
    \includegraphics[width=0.6\linewidth]{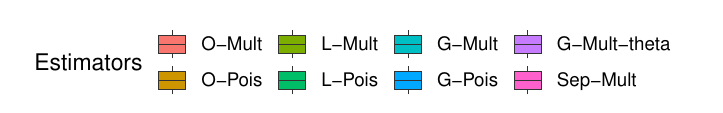}
\caption{Hellinger distances for the competing estimators with $p\in \{10,50\}$ and Scheme $\in \{1,2,3\}$ as $n$ varies.} 
\label{fig:H_D}
\end{figure}

\begin{figure}[tbp]
    \centering
    \begin{minipage}[t]{0.48\linewidth}
    \includegraphics[width=\linewidth]{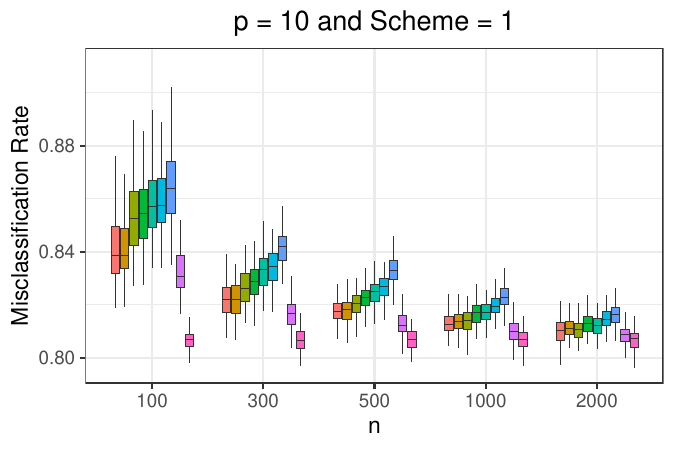}
    \end{minipage}
    \hfill
    \begin{minipage}[t]{0.48\linewidth}
    \includegraphics[width=\linewidth]{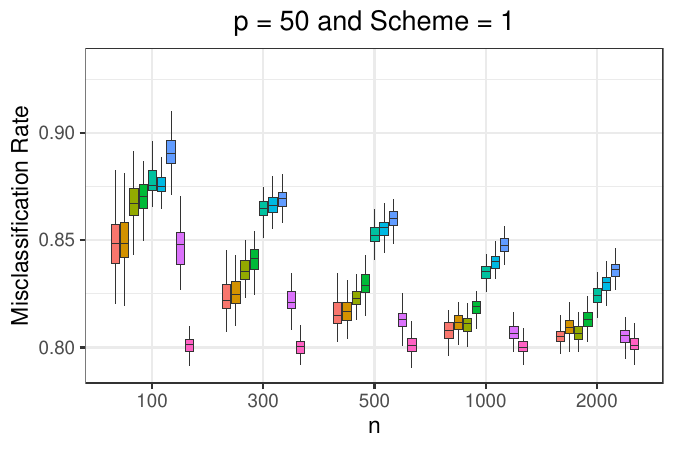}
    \end{minipage}
    
    \begin{minipage}[t]{0.48\linewidth}
    \includegraphics[width=\linewidth]{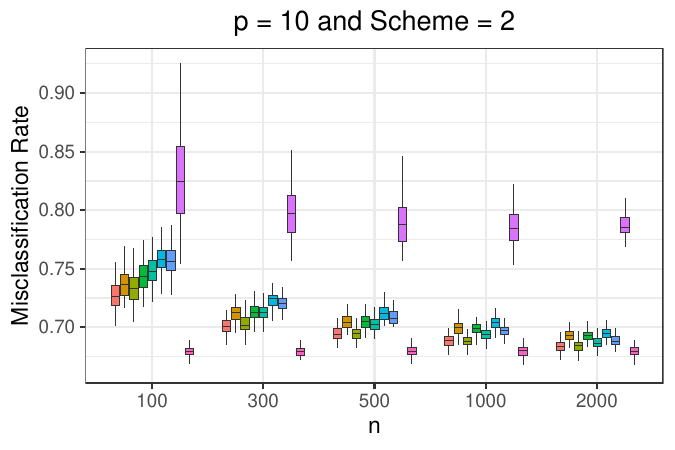}
    \end{minipage}
    \hfill
    \begin{minipage}[t]{0.48\linewidth}
    \includegraphics[width=\linewidth]{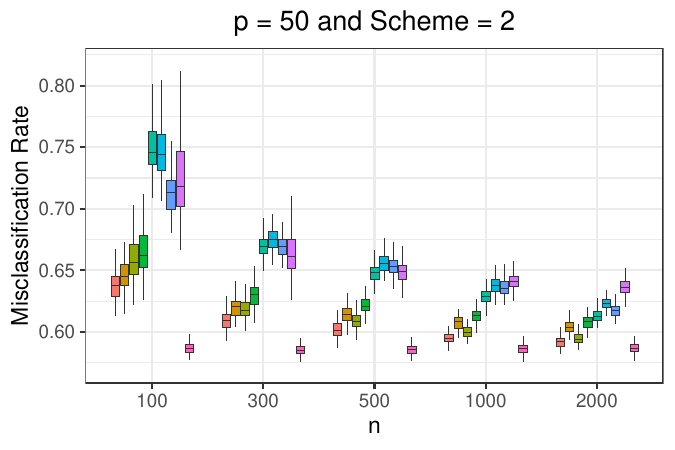}
    \end{minipage}

    \begin{minipage}[t]{0.48\linewidth}
    \includegraphics[width=\linewidth]{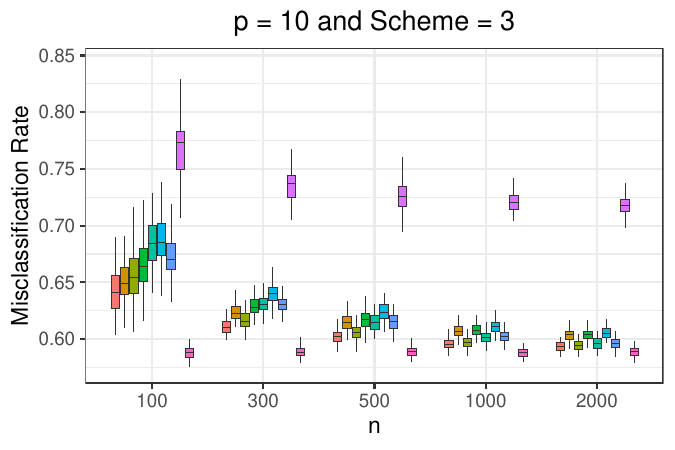}
    \end{minipage}
    \hfill
    \begin{minipage}[t]{0.48\linewidth}
    \includegraphics[width=\linewidth]{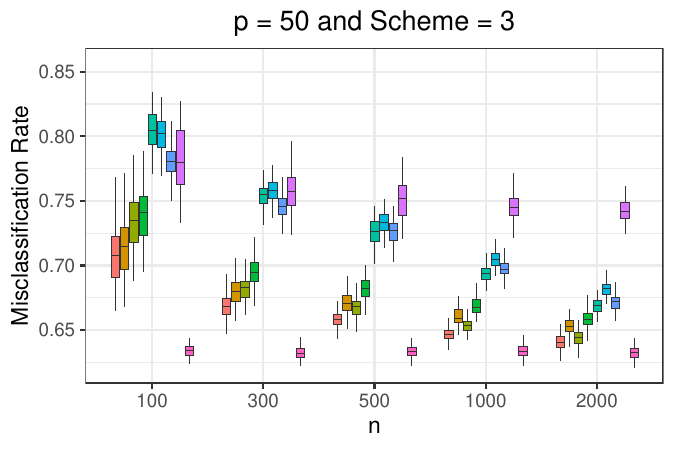}
    \end{minipage}
        \includegraphics[width=0.6\linewidth]{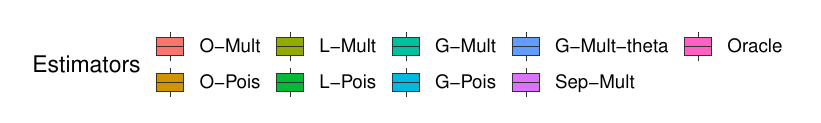}

\caption{Misclassification rates for the competing estimators with $p\in \{10,50\}$ and Scheme $\in \{1,2,3\}$ as $n$ varies.} 
\label{fig:M_R}
\end{figure}

The estimators' performances, evaluated based on Hellinger distance and (joint) misclassification rate on a test set, is displayed in Figures~\ref{fig:H_D} and ~\ref{fig:M_R}. The \textrm{Sep-Mult} estimator is correctly specified under Scheme 1, where the responses are mutually independent. Unsurprisingly, \textrm{Sep-Mult} outperforms all other estimators under this scheme.  Under Schemes 2 and 3 where responses are dependent, we see \textrm{Sep-Mult} perform very poorly relative to the other methods. 

The estimators \textrm{O-Mult}, \textrm{O-Pois}, \textrm{L-Mult}, \textrm{L-Pois}, \textrm{G-Mult}, and \textrm{G-Pois} are all based on our parameterization. Considering the overall performance based on the Hellinger distance and the misclassification rate, the \textrm{O-Mult} estimator is generally the most favorable. This is expected as this method is based on a correct specification of the model and can exploit the hierarchical structure of the local effects. The estimator \textrm{L-Mult} tends to perform second best when sample sizes are large.  Notably, the estimator \textrm{O-Pois} performs reasonably well when $n = 100$: only \textrm{O-Mult} is evidently better. As $n$ increases, however, \textrm{O-Pois} tends to be outperformed by the methods assuming a multinomial data generating model. 

We caution that these results do not imply that estimators based on the multinomial negative log-likelihood are uniformly preferable to those utilizing the Poisson negative log-likelihood. In this case, the multinomial estimators assume a correctly specified model, and thus, as the sample size increases, tend to outperform their Poisson counterparts. 

\subsection{Poisson data generating model}
Simulation study results under the Poisson data generating model are more difficult to interpret than those based the multinomial data generating model. This is partly because when fixing $n$, the effective sample size for the multinomial estimators is a random variable. Specifically, for each of the $n$ samples, we draw a (possibly large) number of Poisson counts from the conditional distribution in \ref{def:poisson_mean}. The multinomial estimators treat each count as an independent realization from a single-trial multinomial. Thus, the number of ``samples'' input into the multinomial estimators can be extremely large and vary greatly from simulation replicate to simulation replicate. For this reason, we exclude results under the Poisson from this manuscript. Nonetheless, to briefly summarize the results we observed in the simulation scenarios we considered (specifically Scheme 2 and 3), we found that under the Poisson data generating models, both L-Pois and G-Pois significantly outperformed L-Mult and G-Mult.

\section{Discussion}
This article introduces an alternative approach to multivariate categorical response regression which relies on an explicit subspace decomposition. Our proposed decomposition allows practitioners to use standard regularization techniques to select the order of effects, and bypasses the issue of dependence on choice of identifability constraints. There are three key directions for future research. 

\subsection{More computationally efficient approaches to hierarchically-structured effect selection}
The use of the overlapping group lasso penalty $\Omega_\mathcal{G}^H$ to select effects adhering to a hierarchy is especially appealing in practice, but leads to an estimator which is more computationally intensive its counterpart excluding hierarchical constraints. In the future, it is important to consider alternative approaches to regularization that may be less computationally intensive, but encourage the desired hierarchy. One such approach may be to utilize the latent overlapping group lasso penalty \citep{obozinski2011group}, which allows the optimization problem to be separable across the (latent) parameters being penalized. This can afford more efficient computational algorithms and schemes to be developed. The estimator based on latent overlapping group lasso penalty is distinct from that based on the overlapping group lasso penalty in the sense that their solution paths are fundamentally distinct, but both can be used to enforce hierarchical constraints. Consequently, the theoretical properties of the estimator based on the latent overlapping group lasso penalty are not immediate from the results we derived in Section \ref{sec:stat_prop}, so this direction is nontrivial.  

Another approach is to use a separable (non-overlapping) approximation to the overlapping group lasso penalty. Specifically, \cite{qi2024non} recently proposed a separable relaxation of the overlapping group lasso penalty, and showed that in terms of squared estimator error,  the estimator using their relaxation is statistically equivalent to that using the overlapping group lasso penalty. Notably, because the relaxation is separable, the corresponding estimator can be computed much more efficiently---roughly at the same cost as estimators using nonoverlapping group lasso penalization schemes.

\subsection{Other representations in predictors}\label{sec:non-parametric}
Recall that in our model \eqref{equ:Poisson_loss} and \eqref{equ:Multivariate_loss}, $\bgtheta \blx=\blH \bgbeta \blx$ is linear in $\blx$. The subspace decomposition model can be extended to accommodate scenarios where the relationship with $\blx$ is not necessarily linear, i.e.,
\begin{equation*}
    \theta (\blx)=\blH \rho (\blx),
\end{equation*}
where $ \theta: \mathbb{R}^p\to \mathbb{R}^{|\blJ|}$ and $\rho:\mathbb{R}^{p} \to\mathbb{R}^{\sum_{s=0}^d L_s }$. Here, \( \rho \) can be associated with both parametric models, such as polynomial regression, and non-parametric models, including splines, kernel-based models, additive models, and deep learning architectures. 

\subsection{Application to the analysis of large contingency tables}\label{sec:non-parametric-app}
Finally, a direction not explored in this article is the use of our estimator for fitting traditional log-linear models for contingency tables. The traditional log-linear model is a special case of our model with the predictor consisting of the intercept only. Effect selection in standard log-linear models has been studied in the past \cite[e.g, see][]{nardi2012log}, but in the asymptotic regime with $n\to\infty$ and all other model dimensions fixed. Thus, it is of particular interest to study whether our finite sample error bounds can be applied, or even refined, in this context.








\end{document}